\documentclass[twocolumn, pre]{revtex4-2}
\ifdefined\added
\else
\usepackage[toc,page]{appendix}
\usepackage{color}
\usepackage{bm}
\usepackage{subfigure}
\usepackage{braket}
\usepackage[pdftex, pdfborder={0 0 0}, colorlinks=true, linkcolor=red, urlcolor=blue]{hyperref}
\usepackage{mathrsfs}
\usepackage{notes2bib}
\usepackage{graphicx,latexsym,setspace,lipsum}
\usepackage{amsmath}
\usepackage{mathtools}
\usepackage{amssymb}
\usepackage{bm}
\usepackage[sort&compress]{cleveref}
\usepackage[usenames,dvipsnames,svgnames,table]{xcolor}
\usepackage{array}
\usepackage{booktabs}
\usepackage[T1]{fontenc}
\usepackage[utf8]{inputenc}
\usepackage{textcomp}
\usepackage{lmodern}
\usepackage{multirow} 
\usepackage{lipsum}
\usepackage{blindtext}
\usepackage{standalone}
\usepackage[normalem]{ulem}
\usepackage{graphicx}
\usepackage{amsfonts}
\usepackage{amssymb}
\usepackage{upgreek}
\usepackage{latexsym}
\fi
\usepackage{mathrsfs}
\usepackage{tikz}
 \ifdefined\added
\else
\newcommand{\added}[2][]{\textcolor{blue}{#2}\textsuperscript{\small\textcolor{red}{#1}}}
\definecolor{shadecolor}{RGB}{80,100,80}
\definecolor{pink}{RGB}{220,100,100}
\usepackage{marginnote}
\newcounter{mparcnt}

\newcommand{\deleted}[1]{}

\DeclareSymbolFontAlphabet{\mathrm}{operators}
\definecolor{red}{rgb}{0.75,0,0}
\definecolor{blue}{rgb}{0,0,0.75}
\definecolor{green}{rgb}{0,0.5,0}
\definecolor{yellow}{rgb}{0,0.5,0}

\def\Eqn#1{Eq.\eqref{#1}}
\def\eqn#1{eq.\eqref{#1}}

\def\tab#1{table~(\ref{#1})}
\def\fig#1{Fig.(\ref{#1})}
\def\app#1{app.~\ref{#1}}

\def\CB#1{ \{ #1 \} }

\def\SBB#1{ \Big[ #1 \Big] }
\def\PB#1{ \Big( #1 \Big)}

\def\comment#1{\iffalse #1 \fi} 
\def\Mcal#1{ \mathcal{#1} }
\def\Mean#1{ \Big\langle #1\Big\rangle }
\def\mbf#1{\mathbf{#1}}
\def\mean#1{\langle #1 \rangle}
\def\T#1{\Tilde{#1}}
\def\mbf#1{\mathbf{#1}}

\def\mcal#1{\mathcal{#1}}
\def\Dmu#1{\Delta\mu#1}

\def\g#1{\gamma#1}
\def\o#1{\omega#1}
\def\r#1{\rho#1}
\def\l#1{\lambda#1}
\def\ddx#1{\mathrm{d^dx}~#1}
\def\ub#1{\upbeta#1}
\newcommand{\eqns}[1]{eqs.(\ref{#1}a-b)}

\def\be{\begin{equation}}
\def\ee{\end{equation}}
\def\bea{\begin{eqnarray}}

\def\a{\alpha}
\def\b{\beta}
\def\p{\partial}
\def\besub{\begin{subequations}}
\def\eesub{\end{subequations}}

\def\P{{\bf P}}

\def\PB#1{ \Big( #1 \Big)}
\def\P#1{( #1 )}
\def\d#1{\nabla#1}
\def\z#1{\zeta#1}
\def\v{{\bf v}}

\def\br{{\bf r}}
\def\G{\Gamma}
\def\d{\delta}

\def\eps{\epsilon}

\def\f{\frac}

\def\bwd{\begin{widetext}}
\def\ewd{\end{widetext}}
\def\dro{\d\r}
\def\dbpi{\d\bm{\pi}}
\def\dpi{\d\pi}
\def\bpi{\bm{\pi}}
\def\k{\mathbf{k}}

\def\ddx{\mathrm{d}^dx~}
\def\rinv{\r^{-1}}

\newcommand{\remark}[1]{}

\newcommand{\postarxiv}[1]{}

\def\bsqr{\scalebox{0.7}{\rotatebox[origin=c]{45}{$\blacksquare$}}}
\newcommand{\Rf}[1]{Ref.~\cite{#1}}
\newcommand{\rf}[1]{Ref.~\cite{#1}}
\newcommand{\Rfs}[1]{Refs.~\cite{#1}}
\newcommand{\rfs}[1]{Refs.~\cite{#1}}
\def\diss{\text{(diss)}}
\def\react{\text{(react)}}
\fi
\def\br{\bm{r}}
\def\hmu{\hat{\mu}}

\newcommand{\ratesym}{\mathord{\tikz[baseline=-0.0ex,scale=0.005]{
      \path[fill=black]
      (15.02245,36.74286)
      .. controls (15.02245,35.88170) and (15.40519,35.11623) .. (16.07498,34.35072)
      .. controls (13.30013,31.38451) and (5.35833,23.25135) .. (0,15.69229)
      -- (1.33958,14.83113)
      .. controls (4.11443,18.65850) and (7.75044,23.82546) .. (17.41456,33.48956)
      .. controls (20.85920,31.67157) and (22.58152,30.61904) .. (22.58152,30.23629)
      .. controls (22.58152,27.84419) and (12.43897,15.59659) .. (12.43897,7.65478)
      .. controls (12.43897,3.63605) and (15.30951,0.00003) .. (19.51962,0.00003)
      .. controls (26.50458,0.00003) and (31.00175,6.69796) .. (37.79534,15.40524)
      -- (36.45576,16.26640)
      .. controls (29.27943,6.60226) and (24.20816,1.81802) .. (21.14625,1.81802)
      .. controls (20.57215,1.81802) and (20.28509,2.10489) .. (20.28509,2.67919)
      .. controls (20.28509,4.97563) and (30.71470,21.24196) .. (30.71470,25.83480)
      .. controls (30.71470,29.37516) and (28.13122,30.14063) .. (18.37141,34.63778)
      .. controls (19.71099,36.64716) and (20.38078,38.08244) .. (20.38078,38.84791)
      .. controls (20.38078,39.99613) and (19.80667,40.57024) .. (18.65846,40.57024)
      .. controls (16.84046,40.57024) and (15.02245,38.46515) .. (15.02245,36.74286)
      -- cycle;
    }}}
 
\begin{document}

\title{Breakdown and Restoration of Hydrodynamics in Dipole-conserving Active Fluids}
\author{Anish Chaudhuri}
\email{anish.chaudhuri@bose.res.in}
\author{Lokrshi Prawar Dadhichi}
\email{lpdadhichi@gmail.com}
\author{Arijit Haldar}
\email{arijit.haldar@bose.res.in}
\altaffiliation{\\All authors contributed equally to this work.}

\affiliation{S. N. Bose National Centre for Basic Sciences, Block JD, Sector III, Salt Lake, Kolkata 700 106, India}

\begin{abstract}
We present a general hydrodynamic theory for active fluids, capable of describing living matter,  that conserve center of mass or dipole moment. Imposition of dipole or center-of-mass conservation has been reported to yield peculiar behavior: breaking Galilean invariance in classical systems and potentially enabling exotic immobile excitations in quantum settings. In passive fluids, dipole conservation has been shown to cause a breakdown of linear hydrodynamics in all experimentally relevant dimensions. We show that introducing activity changes this picture: it can either restore or break linear hydrodynamics depending on 
spatial dimensions. Using our formulation, we 
predict universal dynamical scaling exponents for single-component active fluids in $d=1,2,3$ dimensions and find agreement with microscopic lattice-field simulations. Strikingly, for $d\geq 2$, activity revives linear hydrodynamics, while for $d<2$ it leads to a breakdown;  both cases flow to previously unexplored universality classes. Our results suggest that dipole-conserving active fluids are far more experimentally accessible than their passive counterparts.
\end{abstract}
\maketitle
Constraints are ubiquitous and, by their very nature, often impede the evolution of physical systems\cite{berry2014anomalous,ghosh2015non}. 
In contrast, systems driven into nonequilibrium tend to exhibit signatures of ``enhanced dynamics'', such as super-diffusion and persistent motion capable of overcoming spatial obstacles \cite{chen2016superdiffusion,han2023nonequilibrium}.
From a broader perspective, discovering new phases of matter exhibiting emergent and often unexpected properties is a central pursuit in physics. To this end, the introduction of innovative constraints or nonequilibrium mechanisms provides a promising route to uncover previously unexplored phases.
Particularly, constraints arising from conservation laws, whether emergent\cite{you2020emergent,pretko2017emergent} or otherwise\cite{gromov2020fracton,vijay2016fracton}, are known to produce exotic behavior in matter.
Recently, a new state of matter characterized by slow dynamics and restricted mobility of excitations, dubbed the fracton phase \cite{pretko2018fracton,pretko2020fracton,you2020emergent,grosvenor2021hydrodynamics,gromov2020fracton,osborne2022infinite}, has been discovered. This phenomenon initially examined in quantum systems has been linked to the conservation of moments of either electric charge or mass densities\cite{chamon2005quantum,vijay2015new,pretko2017subdimensional,pretko2017generalized}.The connection between higher-moment conservation and restricted mobility can be illustrated by a simple system conserving both total charge and dipole moment. The motion of an individual unit \deleted{(e.g., a fracton)} violates dipole conservation, whereas only specific correlated motions involving multiple units (e.g., a fracton) are allowed, leading to restricted mobility of individual excitations{\cite{zhai2021fractonic}}. 
Interestingly, conservation of higher moments have also been investigated in the context of quantum lattice-guage and more generally in high-energy physics\cite{nicolis2009galileon,griffin2013multicritical,hinterbichler2014goldstones,griffin2015scalar}.
These developments have generated a huge interest, causing a surge of exploration into the role of higher-moment conservation in classical systems\cite{gromov2020fracton,osborne2022infinite,babbar2025classical,prakash2024classical,prakash2024machian}.\par
More recently, general hydrodynamic descriptions have been put forward to capture the long-wavelength physics of these higher-moment conserving classical fluids under thermodynamic equilibrium\cite{gromov2020fracton,osborne2022infinite,glorioso2022breakdown,guo2022fracton,glodkowski2023hydrodynamics}. 
Hydrodynamics of such dipole- or center-of-mass (COM)-conserving passive fluids have revealed that the resulting constraints hinder fluid motion, altering their typically diffusive behavior into subdiffusive dynamics, along with other striking features. Most notably, as pointed out in Ref.~\cite{glorioso2022breakdown}, COM conservation leads to a breakdown of linear hydrodynamics, causing \deleted{these} passive fluids to become \emph{violently unstable under thermal fluctuations in any experimentally realizable spatial dimension ($d<4$)}.\par
In this article, we study the competition between non-equilibrium dynamics and the kinematic constraint imposed by center-of-mass (dipole) conservation by revealing the fate of active fluids subjected to this constraint.
Active fluids, and in general active matter, describe the physics of living systems and their imitations in the statistical limit\cite{marchetti2013hydrodynamics,
ramaswamy2010mechanics,cates2015motility}.
They are intrinsically out of equilibrium since their constituents consume energy from the surroundings to generate motion
or exert forces, thereby breaking time reversal symmetry at the microscopic level \cite{ramaswamy2017active,dadhichi2018origins,markovich2021thermodynamics}.
The ability of the constituents in active systems to utilize energy gives rise to unusual phases, interfacial phenomena, number fluctuations, organized structures, and other rich behaviors that cannot be realized in their passive counterparts, even when both share the same spatial symmetries\cite{toner1995long,tonertu1998,simha2002hydrodynamic,kumar2014flocking,activeB_NComm,tjhung2018cluster}.\par
Therefore, the important questions that arise, and which we address in this Letter, are: How do active fluids evolve under dipole conservation, and, more importantly, do they exhibit a breakdown of hydrodynamics in all experimentally realizable dimensions, as their passive counterparts \cite{glorioso2022breakdown} do? Answering these questions has significant implications for finding realizations of dipole-conserving active fluids in real-life.\par
To do so, starting from general principles, we develop the hydrodynamics of active fluids with local dipole (COM) conservation and find that activity lowers the critical dimension to $d_c = 2$, implying that linear hydrodynamics is restored for $d > 2$ but breaks down for $d < 2$, in sharp contrast to passive dipole-conserving fluids, where breakdown occurs below $d = 4$\cite{glorioso2022breakdown}.
We verify the critical dimension and predicted scaling exponents by matching them with microscopic lattice-field simulations.
Both the hydrodynamics-restored ($d>2$) and broken ($d<2$) phases of the active fluid, which we uncover, exist solely due to dipole (COM) conservation.
The reason being, dipole conservation allows us to construct a minimalistic formulation of an active fluid whose theory (discussed later) contains a single scalar density field and a momentum-like vector field\deleted{, as discussed later}.
Without the conservation constraint, the minimal momentum-conserving active-fluid theory would require at least a suspension comprising a solute and a solvent\cite{simha2002hydrodynamic}.
Our construction thus provides an example of a hitherto elusive single-component active fluid, highlighting the crucial role of constraints in discovering new phases.\par
The key findings from our construction are as follows:\deleted{The other key results of our work show that} activity enhances the dynamics, as evidenced by the active fluid exhibiting superdiffusive behavior below  $d_c$ and diffusive behavior above it, while the passive fluid under COM conservation remains subdiffusive across all dimensions\cite{glorioso2022breakdown}. Nevertheless, the imprint of COM conservation persists as reflected in the density correlations scaling with wavenumber $k$ as ${\sim k^2}$ in $d\geq d_c$, implying hyperuniformity\cite{hexner2017noise,zheng2024universal,kuroda2023microscopic}, i.e., the vanishing of density fluctuations as the system size diverges. Furthermore, within linear theory, the coupled dispersion relation of the density and longitudinal momentum fields yields either purely diffusive modes $-ik^2$ or sound-like modes $k^2-ik^2$, depending on the regime of phenomenological parameters chosen.

\emph{Hydrodynamic description:}
To construct the hydrodynamics of dipole-conserved active fluids, we begin by identifying the minimal set of slow field variables which dominate in the large length scale and long time limit\cite{chaikin1995principles}. Our choice of the field variables should also ensure that the conserved quantities can be easily expressed in terms of these variables.
The relevant conserved quantities for our construction are total charge (or mass), and the total dipole-moment (COM). In addition, we also demand the conservation of both linear and angular momentum to connect with the notion of conventional fluids such as those described by Navier-Stokes equations.\par
To this end, we find that a scalar density field $\rho$ and a momentum-like vector field $\boldsymbol{\pi}$ are sufficient to meet the above criteria. We refer to $\bm{\pi}$ as a momentum-like quantity rather than momentum since incorporating COM-conservation breaks Galilean invariance \cite{glodkowski2023hydrodynamics} thus making the conventional definition for momentum, $\rho \v$ ($\v$ being velocity), inapplicable. 
Nonetheless, the $\bm{\pi}$ field retains several fundamental properties of the conventional momentum, such as -- sign-change under time reversal, and similar transformations under rotations and translations, thereby making $\bm{\pi}$ a natural substitute for momentum\cite{glodkowski2023hydrodynamics,glorioso2022breakdown}.\par
The conserved quantities expressed in terms of the field variables are
\begin{equation}
    \begin{split}
        Q = \int \ddx \r ~&;~ D_i = \int \ddx x_i \rho \\
        \mathcal{P}_i = \int \ddx \pi_i ~&;~ L_{ij} = \int \ddx (x_i\pi_j - x_j\pi_i)
    \end{split},
    \label{eq:conserved_quantities}
\end{equation}
where, $Q$ is the total mass and $D_{i=1,..d}$ is $i^{th}$ component of centre of mass in $d$ spatial dimensions. The symbol $\mathcal{P}_i$ denotes the components of the total linear momentum obtained from $\pi_i$, the $i$-th component of $\bm{\pi}$, and $L_{ij}$ are the elements of the total angular momentum when expressed in tensorial form. 
Conservation of the above quantities can be 
enforced by choosing the following form of the continuity equations \cite{glodkowski2023hydrodynamics} (\app{supp:Conservation}) \deleted{for mass  $\rho(\mbf{x},t)$ and momentum density field $p_i(\mbf{x},t)$.}
{
\begin{equation}
    \begin{split}
        \p_t \rho + \p_i\p_j J_{ij} = 0 
        ~;~
\p_t \pi_i + \p_j T_{ij} = 0,
        \label{eq:conteqn}
    \end{split}
\end{equation}
where $J_{ij}$, $T_{ij}$ are the currents (fluxes) associated with the $\rho$, and $\bm{\pi}$ fields, respectively. The forms of $J_{ij}$, and $T_{ij}$ will be determined subsequently using the constitutive relations\cite{chaikin1995principles}.} The tensorial nature of the $\rho$-current $J_{ij}$ ensures the conservation of $Q$ and $D_i$ in \eqn{eq:conserved_quantities}. Whereas, the $\bpi$ continuity equation guarantees the conservation of $\mathcal{P}_i$ and setting $T_{ij}$ to be symmetric ($T_{ij}=T_{ji}$) leads to conservation of $L_{ij}$.

To derive the constitutive relations, we follow the approach in \Rfs{marchetti2013hydrodynamics, dadhichi2018origins,markovich2021thermodynamics} and place the dipole-conserving active fluid in contact with a thermal reservoir at temperature $T$. Such a scenario can be described by an entropy production rate $\dot{S}$ related to the free energy $F$ through the relation $T\dot{S}=-dF/dt$.
{The rate of change of free energy $dF/dt$ has an active part (discussed below) and a contribution from the passive free energy density $f$, whose general form is determined by the symmetries and constraints present in the system. In our case, for a fixed temperature $T$, $f$ is a function of the hydrodynamic field $\rho$ and the gradient of the ratio $\nabla (\bm{\pi}/\rho)$. This choice 
ensures dipole-moment (COM) conservation due to the symmetry $f(\rho,\bpi+\r \bm{c})=f(\rho,\bm{\pi})$, where $\bm{c}$ is a constant vector~\cite{osborne2022infinite,glodkowski2023hydrodynamics,glorioso2022breakdown}(\app{app:shift}).}

Activity gets introduced into the rate of free-energy change $dF/dt$ {via an additional term\cite{marchetti2013hydrodynamics,ramaswamy2017active}} representing the energy injected locally into the system per unit time. Such energy injection can arise due to various energy-releasing processes, including chemical reactions that convert ATP molecules to ADP \deleted{molecules} in living systems.  
Considering $\Delta\mu$ to be the energy released per ATP molecule and $\ratesym$ as the rate of molecules consumed per unit volume, 
the entropy production rate for dipole-conserving active fluids (\app{app:force_flux}) is given by 
\begin{equation}
    \begin{split}
        T\dot{S} &= \int d^{d}x \PB{ -\mu \p_t\rho - D_{(ij)} \p_t v_{(ij)} + \ratesym\Dmu },
    \end{split}
    \label{eq:entrop_prod}
\end{equation}
where $ v_i = \rho^{-1} \pi_i,~ v_{(ij)} = \frac{1}{2} (\p_i v_j + \p_j v_i),~D_{(ij)} = {\p f}/{\p v_{(ij)}},~\mu = {\p f}/\p\r$,
and the notation $M_{(ij)}$ denotes the symmetric part of the tensor $M$. 
In general, the presence of activity forces $\Delta\mu>0$, which we consider to be a constant, while $\ratesym$ may vary spatio-temporally.  
Setting $\Delta \mu =0$, in \eqn{eq:entrop_prod} allows us to recover the equilibrium limit of dipole-conserving passive fluids such as those reported in \Rf{glorioso2022breakdown}.

Using the continuity equations (see \eqn{eq:conteqn}), we replace the time derivative of $\rho,\bpi$ in \eqn{eq:entrop_prod} with their respective currents $J_{ij}, T_{ij}$. Simplifying the resulting expression, we obtain a form for $T\dot{S}$ written as a weighted sum of the currents scaled by coefficients (see \app{app:force_flux}). These coefficients are identified as \emph{forces} (and currents as fluxes) within the Onsager-Casimir formalism\cite{marchetti2013hydrodynamics,casimir1945onsager}.\par
We build the constitutive relations by expressing the fluxes ($J_{ij}$, $T_{ij}$) in terms of the forces via the Onsager–Casimir matrix, whose entries are constructed from the hydrodynamic fields ($\rho$, $\bm{\pi}$) and their derivatives (\app{app:onsager_mat}). The dissipative (reactive) components of the fluxes originate from the symmetric (antisymmetric) parts of this matrix. Using the constitutive relations, we replace the currents in the continuity equations (\eqn{eq:conteqn}), thereby closing the system, and obtaining the hydrodynamic equations \deleted{of motion} for the fields  (\app{app:onsager_mat}):
\begin{subequations}\label{eq:EOM}
    \begin{align}
        &\begin{aligned}
\p_t \rho + \a\Dmu \p^2\rho + \ub_1\p^4\mu &+ \ub_2 \Dmu  \p^2\p_l (\rho^{-1} \pi_l)  \\ 
            &+ \ub_3 \Dmu \p_i\p_j (\pi_i \pi_j) + \cdots = 0 \label{eq:EOM_rho}
        \end{aligned} \\
&\begin{aligned}
&\p_t\pi_i - \p_i P + \varsigma_1 \p_i\p^2\mu - \varsigma_2\Dmu \p_i\p_l(\rho^{-1}\pi_l)  + \ell_1\Dmu \p_i \rho \\
&-\varsigma_3\Dmu \p^2(\rho^{-1}\pi_i) + \ell_2\Dmu \p_j (\pi_i\pi_j)  + \cdots = 0 \label{eq:EOM_pi}
        \end{aligned}
    \end{align}
\end{subequations}
where the coefficients $\varsigma_1,\a, \ell_1$ etc. are real phenomenological constants determined by the particular choice of microscopic model {and the ellipsis denote higher order terms}. 
Additionally, the pressure $P$ enters the equations of motion through thermodynamic relations (\app{app:force_flux}), as is typical in fluid dynamics.\par
\emph{Fluctuations:} {Deeper insights into our hydrodynamic construction emerge from a study of}
the fluctuations $\dro$, $\dbpi$ of the $\rho$, $\bpi$ fields, respectively. To obtain the dynamical equations for the fluctuations, we expand \eqns{eq:EOM} around a steady state $\bpi=\bpi_0\equiv 0$ and $\rho=\rho_o$, where $\rho_o>0$ is a finite constant. Retaining terms up to the first non-linear powers of $\dro$, $\dbpi$, we obtain the dynamical equations (\app{app:onsager_mat})
\begin{subequations}\label{eq:nonlin_fluc}
\begin{align}
    &\begin{aligned}[t]
        \p_t \d\r + \a\Dmu \p^2 \d\r + &\b_1 \p^4 \d\r +  \b_2 \p^2\p_l \d\pi_l  \\ 
        + &\b_3\Dmu \p_i\p_j (\dpi_i \dpi_j)= 0 \label{eq:nonlin_eom_rho}
    \end{aligned} \\
&\begin{aligned}[t]
        \p_t \d\pi_i +  \l \p_i \d\r + \z_1 \p_i \p^2 \d\r -& \z_2\Dmu \p_i\p_l \d\pi_l - \z_3\Dmu \p^2 \d\pi_i  \\  
         +& \ell_2\Dmu \p_j(\d\pi_i \d\pi_j) 
        = 0, \label{eq:nonlin_eom_pi}
    \end{aligned}
\end{align}
\end{subequations}
where the new coefficients $\z_1, \l, \cdots$ are obtained from the coefficients in \eqns{eq:EOM}.

Crucially, activity allows diffusive terms, which are 2nd-order in spatial gradients,  to be present in both the $\rho$- and $\bpi$-evolution equations, \deleted{\eqn{eq:nonlin_fluc}(a),(b)} \eqns{eq:nonlin_fluc} respectively. In contrast, these diffusive terms are absent in the passive case \cite{glodkowski2023hydrodynamics,glorioso2022breakdown}\deleted{, where the lowest order gradient appearing in $\rho$ and $\pi$-evolution where fourth order in spatial derivatives}, thereby leading to a subdiffusive behavior dominated by the $\p^4$ terms.
The appearance of the diffusive $\p^2$ terms provide the first indication that activity fundamentally alters the universality class of dipole-conserving fluids.
To uncover these universality classes, and more importantly, to detect the breakdown of linear hydrodynamics, we have retained the lowest order non-linear terms in \eqns{eq:nonlin_fluc}. As shown later, this allows us to find the critical dimension $d_c$ of our construction and identify the regimes in which linear hydrodynamics is restored and broken.
It turns out that even the linear-hydrodynamics--restored regime ($d>d_c$) belongs to a universality class distinct from that of the corresponding passive dipole-conserving fluid\cite{glorioso2022breakdown}.
Therefore, we first discuss the transport properties and scaling exponents of the linearized regime.\par
Linearizing \eqns{eq:nonlin_fluc} and performing a Fourier transform, we get the modes $\dro_\k$, $\dbpi_\k$ in terms of the wave-vector $\k$. Projecting the $\d\bpi_\k$ mode along and perpendicular to $\k$ allows us to decouple the dynamics of the transverse mode ($\dbpi^T = \d\bpi_\k - \hat{\k} . \d\bpi_\k$) from the logitudinal mode ($\d\pi^L = \hat{\mathbf{k}} . \dbpi_k $) and the density mode ($\dro_\k$) (\app{app:lin_th}).
This decoupling leads to a purely diffusive dispersion relation, $\o_{T} = - i \z_3 \Dmu k^2$ for the transverse mode, whose frequency $\o_T$ goes as square of the wavenumber $k\equiv|\k|$, unlike its passive counterpart, where it goes as $-ik^4$ \cite{glorioso2022breakdown}.
The two remaining modes ($e^{i\omega_\pm t}$) are the coupled longitudinal and density modes whose dispersion relations are given by
\begin{equation}
    \begin{split}
\o_{\pm} = \f{1}{2}\PB{- i\G_1 \pm  \sqrt{-\G_2}} k^2,
    \end{split}
    \label{eq:disp_longitudinal}
\end{equation}
where
$\z = \z_2 + \z_3$, $\G_1 = \Dmu(\z - \a)$, and
$\G_2 = 4\l_{}\b_2 + \Dmu^2 (\z+\a)^2$.
Provided parameters are chosen to set $\Im[\o_\pm]<0$, the steady state solution $\bpi=0, \rho=\rho_0$, about which the hydrodynamic equations in \eqns{eq:EOM} were expanded, is guaranteed to be stable. 
Since the transverse mode $\omega_T$ \deleted{(in \eqn{eq:disp_trans})} is purely diffusive, the sign of $\Gamma_2$ determines the existence of sound modes in the theory.

When $\Gamma_2<0$, there are two sound modes whose velocities are given by $\pm\sqrt{-\G_2}k$, respectively. The scaling of velocities with wavenumber $k$ is a typical feature of fractonic systems\cite{glorioso2022breakdown,glodkowski2023hydrodynamics}{, making dipole-conserving fluids distinct from conventional Navier-Stokes fluids where the velocity is independent of $k$}. 
{Nonetheless, \eqn{eq:disp_longitudinal} further indicates that sound modes in dipole-conserving active fluids still share similarities with conventional fluids in that they are damped at a rate proportional to $k^2$.}
Setting $\Gamma_2>0$ turns both sound modes into purely diffusive modes, provided \deleted{$|\Gamma_2|<|\Gamma_1|$} $(\G_1 - \sqrt{|\G_2|}) > 0 $, as required for the stability of the stationary state. This feature provides yet another distinction from conventional fluids, where, in the latter, it is not possible to render all modes diffusive.

It is also interesting to note that analogous dispersion relations appear in the passive dipole-conserving fluid case when energy conservation is explicitly taken into account\cite{glodkowski2023hydrodynamics}, despite the absence of diffusive terms in the $\rho$ and $\bpi$ equations. Similar dispersion relations have also been reported in other works, although these systems are non-fluidic in nature \cite{baggioli2020magnetophonons,delacretaz2022damping,delacretaz2019theory}.\deleted{\Eqn{eq:disp_longi} also indicates that when sound modes are present fluctuations travel [through diffusion as well as sound]//[both diffusively and acoustically]. This can be seen from the amplitude part of the mode, described by $\Im[\omega_\pm]$, which diffuses a distance $\sim \sqrt{\G_1t}$ in time $t$, where as the oscillating part of $\omega_\pm$, given by $\Re[\o_\pm]$, travels a distance $\sqrt{-\G_2}kt$ \emph{ballistically} as sound. 
The fact that both the real and imaginary parts of $\omega$ are quadratic in $k$, in contrast to a non-fractonic fluid, where the real part is proportional to $k$, deserves special comment regarding the diffusive versus ballistic propagation of fluctuations.}
\deleted{In time $t$ a fluctuation diffuses a distance $\sim \sqrt{\G_1t}$, where as it travels distance $\sqrt{-\G_2}kt$ through sound. A fluctuation of wavenumber $k$ travels longer with sound mode for $t>\sqrt{\G_1}/\sqrt{-\G_2}k$ as compared to diffusion and vice versa for $t<\sqrt{\G_1}/\sqrt{-\G_2}k$. In the hydrodynamic limit i.e. $k\rightarrow 0$ limit  this time diverges and hence on measurable time scale fluctuation travel is predominantly diffusive. This is unique to a fractonic system because in normal fluid}
While $\G_2 < 0 ~(\G_2 > 0)$ gives two independent sound (diffusive) modes, setting $\G_2 = 0$ leads to a single mode only. The reason being the existence of an exceptional point \cite{hanai2020critical}, which can be seen by constructing the dynamical matrix of our linear theory (\app{app:excep}). For $\Gamma_2\neq 0$, the two modes are the right eigenvectors of the dynamical matrix having the form $\PB{\f{i k}{2 \bar{\l_{}} } \G^\prime_{-}, 1}^{\mathrm{T}}$, $\PB{\f{i k}{2 \bar{\l_{}} } \G^\prime_{+}, 1}^{\mathrm{T}}$, where $\G^\prime_{\pm}=\Dmu(\a+\z) \pm \sqrt{\G_2}$ and eigenvalues $\o_{\pm}$(see \eqn{eq:disp_longitudinal}). As a side note, we find the angle between the eigenvectors is proportional to the sound speed, $\sqrt{|\Gamma_2|}\,k$ for $\Gamma_2<0$. The exceptional point occurs when $\G_2 = 0$ where the two eigenvectors coalesce into a single diffusive eigenvector $\SBB{(i k/2\l) \Dmu(\a+\z), 1}^{\mathrm{T}}$. Hence, at the exceptional point the rank of the dynamical matrix reduces from two to one. The exceptional point also marks the transition between two dynamical phases, one hosting damped sound modes and the other showing purely diffusive fluctuations.

Introducing white Gaussian noise with zero mean in \eqns{eq:EOM} while preserving the conserved quantities of \eqn{eq:conserved_quantities}  (see \app{app:correlators}), allows us to compute the two-point correlators and the scaling exponents (discussed later). Interestingly, we find, within linear analysis, the equal-time density correlator in all the phases to scale as $k^2$, i.e, $\mean{\d\r(\k,t)\d\r(-\k,t)} \propto k^2$. Such scaling implies hyperuniformity, meaning density fluctuation $\d\r^2$ and compressibility $\kappa = \int d^d x ~\d\r^2$ both vanish as $(k \sim 1/L) \to 0$ and system size $L$ diverges. The resulting number ($N$) fluctuations scale as $\sqrt{\d N^2}\propto\sqrt{V/L^2}\propto L^{\frac{d}{2}-1}\propto N^{\frac{1}{2}-\frac{1}{d}}$, which in 3D gives $N^{1/6}$ \deleted{fluctuations}, much smaller than the equilibrium $N^{1/2}$ scaling. Similar suppression, though perhaps not universal, has been reported in previous studies of dipole-conserving systems\cite{hexner2017noise,zheng2024universal,kuroda2023microscopic}. In our case, the suppression of fluctuations arising from noise occurs because the noise contribution enters at higher order in the wavenumber due to dipole conservation.\par
\emph{Scaling exponents \& Critical dimension:}
\begin{table}
    \centering

    \resizebox{0.825\columnwidth}{!}{\setlength{\arrayrulewidth}{0.3pt}

    \begin{tabular}{ |c|c|c| }
    \cline{2-3}
    \multicolumn{1}{c|}{}& {\bf active} & {\bf passive}\\
    \hline
    $d_c$ & 2 & 4\\
    \hline
    $z\ (d<d_c)$ & $1+d/2$ & $2+d/2$\\
    \hline
    $z\ (d\geq d_c)$ & $2$ & $4$\\
    \hline
    \end{tabular}
    }

\caption{Predictions for the critical dimension $d_c$ and the dynamical exponent $z$ across dimensions $d$ for dipole-conserving active fluids vs their passive counterparts. }
    \label{tab:Tabulated_z}
\end{table}
Our construction readily allows us to predict the critical dimension $d_c$ above (below) which linear hydrodynamics is restored (broken). Rescaling $ \mbf{x} \to b \mbf{x}, ~t \to b^z t, ~\d\r \to b^\varrho \d\r , ~\d\bpi \to b^\Upsilon \d\bpi$ in 
\eqns{eq:nonlin_fluc} {in the presence of Gaussian white noise (see \app{app:onsager_mat})} by the system scaling parameter $b>1$, and dividing by the resulting coefficient of the term $\partial_t$\cite{forster1977large} gives
\begin{subequations}
    \begin{align}
        &\begin{aligned}
            &\p_t \d\r + b^{z-2} \a\Dmu \p^2 \d\r + b^{(z - 3 - \varrho + \Upsilon)} \b_2 \p^2\p_l \d\pi_l \\
&+ b^{z-4} \b_1 \p^4 \d\r
            {+ b^{(z - \varrho - 2 + 2 \Upsilon)}\b_3\Delta\mu \p_i\p_j (\dpi_i \dpi_j)} \\
            &= b^{(\frac{z}{2} - \varrho - \frac{d}{2} - 2)} \p_i\p_j\psi_{ij} 
            \label{eq:rho_scaled}
        \end{aligned}\\
&\begin{aligned}
            &\p_t \d\pi_i + b^{(z + \varrho - \Upsilon - 1)} \lambda\p_i \d\r - b^{(z-2)}  \z_3\p^2 \d\pi_i \\
&+ b^{(z + \Upsilon - 1)} \l_2 \Dmu \p_j(\d\pi_i \d\pi_j)  
            = b^{(\frac{z}{2} - \Upsilon - \frac{d}{2} - 1)} \p_j \eta_{ij}, 
            \label{eq:pi_scaled}
        \end{aligned}        
    \end{align}\label{eq:scaled_eom}
\end{subequations}
where $\varrho$, $\Upsilon$ are the scaling exponents for the $\rho$, $\bpi$ fields, and $z$ is the \emph{dynamical exponent}. 
he noise correlator for $\eta_{ij}$ satisfies $\mean{\eta_{ij}(\mbf{x}, t) \eta_{lm}(\mbf{x}^\prime, t^\prime)}=L_{ijlm}\d(\mbf{x} - \mbf{x}^\prime) \d(t - t^\prime)$, where $L_{ijlm}$ is a tensor symmetric under the exchange of $i\leftrightarrow j$, $l\leftrightarrow m$ and, due to nonequilibrium, not dictated by the fluctuation dissipation theorem. A similar relation holds for $\psi_{ij}$ as well. 

For detecting the linear hydrodynamics regime, we demand the scaling exponents of the coefficients corresponding to the linear terms and noise fields ($\psi_{ij}$, $\eta_{ij}$) in \eqns{eq:scaled_eom} to be \emph{zero}.
This choice leads to $z = 2, ~\varrho = -(1 + \frac{d}{2}), ~\Upsilon = - \frac{d}{2}$, ensuring that the hydrodynamic equations remain scale-invariant when the non-linear terms are treated absent. Therefore, the above values for the scaling exponents define the Gaussian fixed point of our theory.
The non-linear terms, when included back, scale differently than the linear terms at this fixed point. In particular, the fastest growing nonlinear terms $\p_i\p_j (\dpi_i \dpi_j)$ and $\p_j(\d\pi_i \d\pi_j)$, in \eqns{eq:scaled_eom}, scale with the exponent $\xi\equiv z - \varrho - 2 + 2 \Upsilon=z + \Upsilon - 1=1-d/2$.

When $d<2$, $\xi>0$, implying that the non-linear term grows under system scaling and destabilizes the Gaussian fixed point, whereas non-linearities become either irrelevant or marginal for $d\geq2$, since $\xi\leq 0$, and the fixed point is stable. Consequently, the critical dimension for dipole-conserving active fluids is $d_c=2$.
The analytical values of $d_c$ and the dynamical exponent $z=2$ for $d\geq d_c$ offer concrete predictions that can be tested in simulations (discussed next) and, more importantly, in experiments.

While a prediction for $z$ when $d<d_c$ requires a formal and elaborate RG analysis\cite{forster1977large}, an approximate value can be obtained by taking clues from the instability of the Gaussian fixed point. 
Inspired by \cite{glorioso2022breakdown} we make a \emph{crude} approximation of $z$ by 
considering the time-derivative to scale similarly
with the non-linear term responsible for the Gaussian fixed point instability, i.e. $\p_t \d\bpi\approx\p_i \d\bpi^2 $ and assuming that the scaling of $\d\r$, $\d\bpi$ do not renormalize significantly even in the presence of relevant non-linearities. This dimensional balancing gives $b^{-{(z+d/2)}}\approx b^{-(d+1)}$ leading to $z=1+d/2$ when $d<d_c$(\app{app:scaling}).

We summarize the above analytical predictions for $z$ in the hydrodynamics-broken and restored regimes, and compare them with their passive-fluid counterparts \cite{glorioso2022breakdown}, in \tab{tab:Tabulated_z}, highlighting the role of activity in modifying these predictions.\par
\emph{Simulations:}
To test the predictions for $z$ and $d_c$ (\tab{tab:Tabulated_z}), we utilize a microscopic lattice model defined on the $d$-dim cubic lattice, comprising point particles of mass $m=1$ residing on each lattice site and interacting with their nearest neighbors (nn).
Since hydrodynamics is insensitive to microscopic details, any lattice Hamiltonian that conserves the quantities in \eqn{eq:conserved_quantities} should suffice to generate the passive sector of the long-wavelength dynamics for our hydrodynamic theory. Hence, the following Hamiltonian, introduced in \rf{glorioso2022breakdown}, will also apply to active fluids:
\begin{align}
    &H = \sum_{\br} \left[
        \sum_{\mu \nu} \frac{K_\pi}{2}(p^{(\mu)}_{\br} - p^{(\mu)}_{\br + \hat{\nu}} )^2  + \sum_{\mu} V(x^{(\mu)}_{\br} - x^{(\mu)}_{\br + \hat{\mu}}) \right] \nonumber \\
        &V(x) = \frac{1}{2} K_2 x^2 + \frac{1}{3} K_3 x^3 + \frac{1}{4} K_4 x^4, \label{eq:latt_model_ddim}
\end{align}
where $x^{(\mu)}_{\br}$, $p^{(\mu)}_{\br}$ are the $\mu$-direction components of displacement and conjugate momentum of the $\br$-th site particle, and $V(x)$ represents the nn interactions controlled by the strengths $K_{2,3,4}$.
The coefficient $K_\pi$ represents a momentum stiffness (see \app{app:latt_model}) encoding an energetic cost for momentum gradients and thereby leading to dipole conservation.
The above form of $H$ ensures conservation of $D^{(\mu)}=\sum_{\br} m x^{(\mu)}_{\br}$, $P^{(\mu)} = \sum_r p^{(\mu)}_{\br}$, and $Q = \sum_{\br} m$, since their respective Poisson brackets with $H$ vanish (see \app{app:latt_model}).

We now deviate from equilibrium (and \rf{glorioso2022breakdown}) to introduce activity into our lattice model by -- \emph{first}, explicitly incorporating dissipation \ $\gamma$   and temporal-noise $\zeta^{(\mu)}_{\br}(t)$, respecting $D^{(\mu)}$, $P^{(\mu)}$, and $Q$ conservation, in the dynamical equation for $p^{(\mu)}_{\br}$ generated by $H$:
\begin{equation}
    \begin{split}
\dot{p}^{(\mu)}_{\br} &= - \frac{\d H}{\d x^{(\mu)}_{\br}} - \g_a \dot{x}^{(\mu)}_{\br} - \z^{(\mu)}_{\br} + \z^{(\mu)}_{\br - \hmu}. 
\end{split}
    \label{eq:stochastic_EOM}
\end{equation}
{The above choice of noise and dissipation terms further ensures that the fluctuation-dissipation theorem (FDT) is inherently broken (\app{app:FDT}) {unlike \cite{glorioso2022breakdown}} and pushes our model out of equilibrium at the microscopic level.}
\emph{Second}, we choose the noise to have zero mean, $\mean{\zeta^{(\mu)}_{\br}}=0$ and fix its correlator to satisfy
$\mean{\z^{(\mu)}(\br,t) \z^{(\nu)}(\br^\prime,t^\prime)} = 2 T \d_{\mu \nu} \d_{\br \br^\prime} \d(t-t^\prime)$,
{so as to mimic the effect of a thermal bath maintained at temperature $T$.}

\begin{figure}
    \centering
\includegraphics[width=\columnwidth]{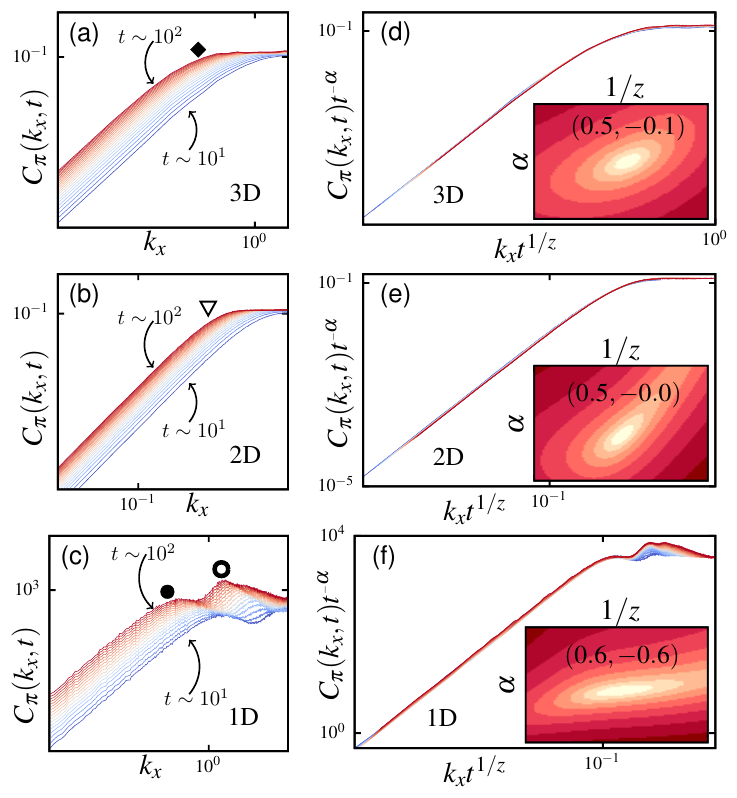}
    \caption{Panels (a--c) show log–log plots of the momentum correlator
    $C_\pi(k_x,t)= \mean{p^{(x)}_{k_x, k_\perp}(t)p^{(x)}_{-k_x, -k_\perp}(t)}|_{k_\perp=0}$ versus wavenumber $k_x$ at different times $t$
    in dimensions $d=3,2,1$, respectively. Panels (d--f) show the scaled forms
    $C_\pi(k_x t^{1/z}) t^{-\alpha}$, corresponding to panels (a--c), and their collapse onto a universal curve. Insets display the respective optimal $\a$ and $z$ values, obtained by minimizing the collapse error in the $(\a, z^{-1})$ plane (see \app{app:collapse_err}). Independent confirmation of optimized $z$ values is obtained by tracking the time evolution of the marked features ($\bsqr,~\bm{\triangledown},~\circ,~\bullet$).(See \app{app:tvsk} for details.)}\label{fig:collapse}
\end{figure}

To extract $z$ and $d_c$, we examine the equal-time correlator
$C_\pi(k_x,t)= \mean{p^{(x)}_{k_x, k_\perp}(t)p^{(x)}_{-k_x, -k_\perp}(t)}|_{k_\perp=0}$,
{setting the transverse component $k_\perp=0$}, which should attain the universal scaling form $C_\pi(k_x,t)=t^{\alpha}F(k_x/t^{1/z})$ if our lattice model reaches the hydrodynamic limit. Furthermore, for $d \geq d_c = 2$, linear hydrodynamics at the Gaussian-fixed point predicts $C_\pi(k_x,t \to \infty) \sim k_x^0$, even in the presence of nonlinearities $K_{3,4} > 0$ in $V(x)$. Provided our simulations confirm the predictions summarized in \tab{tab:Tabulated_z} and exhibit the above features, we can consider our hydrodynamic description to be successful.

We plot $C_\pi(k_x,t)$ from simulations at various times $t$
for dimensions $d=3,2,1$ (top to bottom) in \fig{fig:collapse} (left column);
and immediately observe its behavior in $d=1$ to differ qualitatively from that in $d=2,3$. We attempt to collapse $C_\pi(k_x, t)$ onto a scaling form by numerically searching for optimal values in the $\alpha$--$z^{-1}$ plane and indeed find that $t^{-\alpha} C_\pi(k_x/t^{1/z})$ collapses onto a single curve for all dimensions $d=1,2,3$, as shown in \fig{fig:collapse} (right column), implying the hydrodynamic description to be valid. Moreover, after optimization, we obtain $z=2$ for $d=2,3$ and
$z=1.66\ldots \approx 1 + d/2$ for $d=1$, in close agreement with the
predictions in \tab{tab:Tabulated_z} and thus confirming the critical dimension
$d_c=2$. The long-time ($t\sim10^2$) curves in \fig{fig:collapse}(a,b) further show $C_\pi(k_x,t)$ becoming independent of $k_x$, consistent with the prediction $C_\pi(k_x,t) \sim k_x^0$, again indicating that the Gaussian fixed point holds for $d \geq d_c$ and breaks down for $d < d_c$ in the linear-hydrodynamics--broken regime.

\emph{Discussion:}
The close agreement between simulations and hydrodynamic theory establishes hydrodynamics-broken and hydrodynamics-restored phases as distinct universality classes of dipole-conserving active fluids. Thus, answering the question--what is the fate of active fluids under dipole conservation? Both phases, previously unreported, are distinct from their passive and Navier–Stokes counterparts: in the hydrodynamics-restored phase, this distinction appears in the linear dispersion and nature of the sound modes, while in the broken phase, it manifests through a distinct dynamical exponent.

We identify the upper critical dimension for dipole-conserving single-component active fluids as $d_c=2$, in contrast to $d_c=4$ for passive fluids \cite{glorioso2022breakdown}. Linear theory is therefore expected to be quantitatively accurate in three dimensions and in two dimensions up to logarithmic corrections. These predictions can be directly tested, particularly in 2d, using laser-activated Janus particles with optically controlled dynamics \cite{walther2013janus}.

Our work initiates the exploration of dipole, or center of mass, conservation in active matter.
Natural extensions include the impact of dipole conservation on active nematics\cite{ramaswamy2003active,zhai2021fractonic} and multicomponent active fluids.
\\
\\
\\
\emph{Acknowledgments:}
A.C. thanks Debraj Dutta for helpful discussions regarding numerical methods.  
\clearpage
\begin{appendix}
\def\MAIN{}
\clearpage{}\section{Continuity equation}\label{supp:Conservation}
We provide a brief derivation of the continuity equation required for both dipole (center-of-mass) conservation and charge conservation. First, the conservation of the total charge, $Q = \int \ddx \rho(\mathbf{x},t)$,  implies
\begin{equation}
\partial_t\rho(\bm{x},t)=-\nabla\cdot {\mathbf J},
\label{eq:charge_cont}
\end{equation}
where ${\mathbf J}=\sum_{i=1}^d J_i \hat{\mathbf{e}}_i$ is the corresponding charge (mass) current density.
Next, dipole conservation leads to the global conserved quantity $D_i=\int \ddx \rho(x)x_i$, such that $\partial_tD_i=0$, which mathematically implies
\begin{align}
\partial_t D_i=& \int \ddx x_i\partial_t\rho(x) 
=
-\int \ddx x_i\nabla\cdot {\mathbf J}\notag\\
=&-\oint_{\p \Omega} (d\mathbf{S}\cdot\mathbf{J})x_i + 
\int \ddx J_i.\label{eq:entpro}
\end{align}
The first term in the above expression is a boundary term and vanishes under appropriate boundary conditions. Therefore, for dipole conservation, the contribution of the last integral in \eqn{eq:entpro} must vanish in a similar manner, which happens if $\int \ddx J_i$ is also a surface integral; meaning $J_i$ takes the form $J_i=\partial_jJ_{ij}$ described by the 2nd-rank tensor $J_{ij}$. Expressing ${\mathbf J}$ using the said form in \eqn{eq:charge_cont} gives the continuity equation
\begin{equation}
\partial_t\rho=-\partial_i\partial_jJ_{ij}
\end{equation}
reported in the main text.
\section{Free energy form}
\label{app:shift}
As shown in \rf{glodkowski2023hydrodynamics}, for a dipole-conserving fluid, the internal energy density $\epsilon$ is a function of the entropy density $s$, $\rho$, and $\p_i(\rinv\pi_j)$. The internal energy density is related to the free energy density $f$ by the standard Legendre transform, $f = \epsilon - Ts$, making $f$ a function of $\rho$, $\p_i(\rinv\pi_j)$, and $T$. Since $T$ is kept constant in our system by an attached thermal bath, $f$ is a function of $\rho$ and $\p_i(\rinv\pi_j)$.
Another way to reach the same conclusion is to realize that the free energy is equivalent to the coarse-grained Hamiltonian \cite{osborne2022infinite,chaikin1995principles}, and only this form of the free energy will produce the reactive terms (from the coarse-grained Poisson bracket) \cite{osborne2022infinite} that conserve dipole in the hydrodynamic equation for $\r$.
\section{Identification of forces and fluxes}\label{app:force_flux}
Derivation of the generalized hydrodynamic equations follows from the identification of fluxes and forces contributing to the entropy production rate. 
Keeping the temperature $T$ constant by coupling the system to a heat reservoir, the entropy production rate $\dot{S}$ determined by the free energy $F$ of the active dipole-conserving fluid as \cite{marchetti2013hydrodynamics}. 
\begin{equation}
T\dot{S}=-dF/dt.
\label{eq:entropy_prod}
\end{equation}\par
As stated in the main text, the free energy density consists of passive and active contributions. In \app{app:shift} we have justified that the passive free energy density $f$ is a function of $\r$ and $\p_i(\rinv \pi_j)$. Furthermore, as shown in Ref.~\cite{glodkowski2023hydrodynamics}, for a system preserving parity $f$ must depend only on the symmetric part of  $\p_i(\rinv \pi_j)$.
Therefore, its differential reads
\begin{equation}
    \begin{split}
    d f = \mu d\rho + D_{(ij)} d v_{(ij)} ~&;~ v_{(ij)} = \f{1}{2}(\p_i v_j + \p_j v_i) \\
    v_j &= \r^{-1}\pi_j~.
    \end{split}
    \label{eqapp:df}
\end{equation}
In \eqn{eqapp:df}, \(\mu\) is the chemical potential conjugate to the charge density \(\r\), while \(D_{(ij)}\) denotes the variable conjugate to \(v_{(ij)}\). Hence,
\begin{equation}
    \mu = \frac{\p f}{\p\r_{}},~D_{(ij)} = \frac{\p f}{\p v_{(ij)}}~.
    \label{eqapp:mu_Dij}
\end{equation}
Furthermore, the leading order form of $D_{(ij)}$ can be derived. Due to symmetries and constraints discussed in \app{app:shift}, the passive free energy density can be expanded as $f(\r,v_{(ij)}) = \sum_{u = 0,1,\cdots} \f{c_{u}(\r)}{u}(v_{(ij)})^u$. Hence, using the definition from  \eqn{eqapp:mu_Dij} one finds $D_{ij} = \sum_u c_u(\r)(v_{(ij)})^{u-1}$. 
\postarxiv{Why not place this under \app{app:shift} or integrate \app{app:shift} under \eqn{eq:entropy_prod}?}
\postarxiv{\\---------------\\}

Keeping upto the leading order in $v_{(ij)}$, we obtain  
\begin{equation}
    D_{(ij)} = \r v_{(ij)}~.
    \label{eqapp:Dij}
\end{equation}\par
Activity enters as a local energy-injection term\cite{marchetti2013hydrodynamics,ramaswamy2017active} in $dF/dt$, arising from energy-releasing processes, such as ATP-consuming chemical reactions characterized by a release rate ($\ratesym$) and the energy released per ATP molecule ($\Dmu$).
Maintaining $\Delta\mu$ at a constant value effectively amounts to a continuous supply of energy, turning the system active. The corresponding rate of change of the free energy per unit volume is then $-\ratesym\Delta\mu$. Accounting for both passive ($-\partial_t f$) and active ($\ratesym\Delta\mu$) contributions, the entropy production rate, Eq.~\eqref{eq:entropy_prod}, for dipole-conserving active fluids is given by
\begin{equation}
    T\dot{S} = \int d^d x \PB{ -\mu \p_t\rho - D_{(ij)} \p_t v_{(ij)} + \ratesym\Dmu }
\label{eqapp:TSdot}
\end{equation}
as reported in the main text.

To express the entropy production in the standard flux–force form, we eliminate the time derivatives of the conserved variables by substituting $\p_t\r$ and $\p_t\bpi$ from \eqn{eq:conteqn} in the main text, and performing integration by parts (neglecting the boundary terms), the integrand of \eqn{eqapp:TSdot} can be expressed as a weighted sum of currents $(J_{ij},T_{ij},\cdots)$ as
\begin{equation}
    \begin{split}
        T \dot{S} = \int d^dx ~\PB{J_{(lm)} A_{(lm)} + T_{(kj)} W_{(kj)} + \ratesym \Dmu}
    \end{split},
    \label{eqapp:forces_times_fluxes}
\end{equation}
where 
\begin{equation}
    \begin{split}
        A_{(lm)} = \p_l\p_m \Lambda&,~ \Lambda = \mu + (\rho^{-1}\pi_j)W_j,\\
        W_{(kj)} = \p_{(k} W_{j)} = \f{1}{2}(\p_k W_j &+ \p_j W_k),~W_j = \rho^{-1} \p_i D_{(ij)}
    \end{split}
\end{equation}

We now introduce the thermodynamic pressure. Following a similar approach to \rfs{glorioso2022breakdown,glodkowski2023hydrodynamics}, we add a vanishing contribution $- \p_i (PW_i) + P \p_i W_i + W_i\p_iP = 0$ to \eqn{eqapp:forces_times_fluxes}, which results in
\begin{equation}
    \begin{split}
        T \dot{S} =  \int d^dx ~\PB{ (T_{(kj)} &+ P \delta_{kj}) W_{(kj)} + \\ 
        &J_{(lm)} A_{(lm)} + \ratesym \Dmu  + W_i \p_i P }.
    \end{split}
    \label{eq:TSdot_with_P}
\end{equation}
Using standard thermodynamic relations\cite{wittkowski2014scalar} for pressure $P$ given by $P = \mu\rho - f ~;~ dP = \r d\mu - D_{(ij)} d v_{(ij)}$, one can write $\p_i P = \rho \p_i \mu - D_{(lm)} \p_i\p_{l}(\rho^{-1}\pi_m)$ and substitute this expression for $\p_i P$ in \eqn{eq:TSdot_with_P} to yield
\begin{equation}
        \begin{split}
            &T \dot{S} =  \int d^dx ~\PB{(J_{(lm)} - D_{(lm)}) A_{(lm)} + \ratesym \Dmu \\ 
            &+ \P{T_{(kj)} + P \delta_{kj}} W_{(kj)} + W_{ij} \PB{D_{im}\p_j v_m - \p_l D_{(il)} v_j}}.
        \end{split}
        \label{eqapp:last_term}
\end{equation}
Next, we express the above integrand as a weighted sum of symmetric tensors. To this end, we show that the combination $W_{ij} \PB{D_{im}\p_j v_m - \p_l D_{(il)} v_j}$ contains only symmetric contributions. To demonstrate this explicitly, we introduce an additional identity
\begin{equation}
    \begin{split}
    \p_i \PB{W_j \p_k \PB{D_{(ij)} v_k &- D_{(kj)} v_i} } \\
    &- \PB{\p_k \PB{D_{(ij)} v_k - D_{(kj)} v_i}} \p_i W_j = 0
    \end{split}
    \nonumber
\end{equation}
in \eqn{eqapp:last_term}. After algebraic simplification, we obtain
\begin{equation}
    \begin{split}
        &T \dot{S} = \int \ddx ~(J_{(lm)} - D_{(lm)}) A_{(lm)} + \P{T_{(kj)} + P \delta_{kj}} W_{(kj)} \\ 
        &+ \ratesym \Dmu + \PB{ \p_k\PB{D_{(ij)} v_k} - W_i v_j\rho - W_j v_i \rho } W_{(ij)} + \mathcal{T}_{ij} W_{ij},
    \end{split}
    \label{eq:the_term}
\end{equation}
where
\begin{equation}
    \mcal{T}_{ij} = \PB{ D_{(im)}\p_j v_m - D_{(jk)}\p_k v_i }.
\end{equation}
The term $\mathcal{T}_{ij}$ in \eqn{eq:the_term} is symmetric, since its antisymmetric part vanishes by rotational invariance of the free energy. To show this, we consider a constant antisymmetric tensor $\Theta_{[ij]}$ contracted with $\mathcal{T}_{ij}$ and integrated over space  
\begin{equation}
    \begin{split}
        \int \ddx \mathcal{T}_{ij} \Theta_{[ij]} 
        = \int \ddx D_{(im)} \PB{\p_j v_m \Theta_{[ij]} + \p_i v_c\Theta_{[mc]}}.
    \end{split}
    \nonumber
\end{equation}
Since under an infinitesimal rotation $x^\prime_i = x_i + \eps \Theta_{[ij]}x_j,~$ variation of $\p_i v_m$ is $\d (\p_i v_m) = \PB{\p_i v_c \Theta_{[m c]} + \p_j v_m\Theta_{[ij]}}$, hence,
\begin{equation}
    \begin{split}
\int \ddx \mathcal{T}_{ij} \Theta_{[ij]} = \int \frac{\p f}{\p (v_{(im)})} \d(v_{(im)}) = \delta F = 0.
    \end{split}
    \nonumber
\end{equation}
The expression vanishes due to the rotational invariance of the free energy. Therefore, we can conclude $\mathcal{T}_{ij} W_{ij} = \mathcal{T}_{ij} W_{(ij)}$. Consequently,
\begin{equation}
    \begin{split}
        T \dot{S} = \int d^d x \PB{A_{(lm)} J^{\prime}_{(lm)} + W_{(ij)} T^\prime_{(ij)} + \ratesym \Dmu},
    \end{split}
    \label{eqapp:entroprod_force_flux}
\end{equation}
where, 
\begin{equation}
    \begin{split}
        J^\prime_{(lm)} = J_{(lm)} - D_{(lm)} ~&;~ D_{(lm)} = \frac{\p f}{\p v_{(lm)}}  \\ 
        \mcal{T}_{ij} = D_{(i\a)} \p_j v_\a - D_{(kj)} \p_k v_i ~&;~ 
T^\prime_{(ij)} = T_{(ij)} + P \d_{ij} + \Xi_{(ij)} 
        \\
        \Xi_{(ij)} = \p_k\PB{D_{(ij)} v_k} &- ( W_{(i} v_{j)} )\r + \Mcal{T}_{(ij)}
    \end{split}
    \label{eqapp:effective_curr}
\end{equation}
Finally, \eqn{eqapp:entroprod_force_flux} expresses the rate of entropy production in a familiar flux–force structure. This form makes the pairing between thermodynamic forces and their conjugate fluxes explicit, thus allowing for a systematic identification of the independent contributions. The resulting force–flux pairs are listed below.
\begin{table}[h!]
\centering
\setlength{\arrayrulewidth}{0.3pt}

\begin{tabular}{|c|c|c|}
\hline
\textbf{Flux}            & \textbf{Force}           & \textbf{TRS}              \\ \hline 
$J^\prime_{ij}$                 &  $ A_{ij}$              & +1                         \\ \hline 
$T^\prime_{ij}$               &  $ W_{ij}$              & -1                         \\ \hline
$\ratesym$                      &  $\Dmu$                  & +1                         \\ \hline
\end{tabular}
\caption{Fluxes and their corresponding forces, together with their time-reversal signatures. These quantities govern entropy production, and their corresponding contributions to entropy production must be non-negative. For this reason, each flux and its associated force share the same time-reversal signature.}
\label{tab:flux_force_TRS}
\end{table}

\section{Hydrodynamic equations of motion}\label{app:onsager_mat}
Within the framework of hydrodynamics, the constitutive equations of an active system are formulated by establishing the most general linear relations between fluxes and forces that are compatible with the underlying symmetries of the system \cite{marchetti2013hydrodynamics,dadhichi2018origins,markovich2021thermodynamics}. The tabulated (\tab{tab:flux_force_TRS}) forces and fluxes are coupled through the Onsager–Casimir matrix \cite{casimir1945onsager}, whose elements are constructed from hydrodynamic variables and their gradients. The entries of this matrix may be scalars, vectors, or second-rank tensors. To retain the most relevant contributions, we restrict the entries to be of the lowest order in gradients and fields. The scalar density $\rho$ and the momentum density $\bpi$ are the natural candidates for scalar and vector entries, respectively. Second-rank tensors are constructed from the hydrodynamic variables $\rho$, $\bpi$, and their gradients, whose multiple canddidates are 
\begin{equation}
    \begin{split}
        q_{(ij)} = \pi_i\pi_j ~,~ Q_{(ij)} &= 1/2(\pi_i \p_j\r + \pi_j\p_i\r)\\
        \tau_{(ij)} = \rho \delta_{ij} + q_{(ij)} + R_{(ij)} ~&,~ \Omega_{(ij)} = Q_{(ij)} + \chi_{(ij)} \\
        R_{(ij)} &= (\p_i\rho)(\p_j\rho) .
    \end{split}
    \label{eq:compiled_tensors}
\end{equation} 
A key aspect is the role of time-reversal symmetry: each flux must be decomposed into two distinct contributions. The reactive part has a time-reversal signature opposite to that of the product of its conjugate force and the corresponding matrix element, whereas the dissipative part has the same time-reversal signature as this product. \\ 
For analytical convenience, all second-rank tensors are decomposed into their isotropic (trace) and traceless components according to $\Sigma_{ij} = (\Sigma/d) \d_{ij} + \T{\Sigma}_{ij}$, where $\Sigma \d_{ij}$ denotes the diagonal isotropic part and $\T{\Sigma}_{ij}$ represents the traceless part. Furthermore, all fluxes are decomposed into their reactive and dissipative parts as follows
\begin{equation}
    \begin{split}
        J^{\prime}_{ij} = J^{\prime \diss}_{ij} + J^{\prime \react}_{ij} \\ 
        T^{\prime}_{ij} = T^{\prime \diss}_{ij} + T^{\prime \react}_{ij} \\ 
        \ratesym = \ratesym^\diss + \ratesym^\react.
    \end{split}
\end{equation}

\emph{\textbf{Dissipative fluxes:}} The dissipative Onsager matrix is symmetric and couples forces and fluxes such that the resulting product of the force and the corresponding matrix element has the same time-reversal signature as the flux. This leads to the following set of equations:
\begin{equation}
    \begin{split}
         J^{\prime \diss} &= \a_{10} A + \a_{12} \tau \Dmu 
        \\
        \T{J}^{\prime \diss}_{(ij)} &= \a_{13} \T{A}_{(ij)} + \a_{14} (\T{q}_{ij} + \T{R}_{ij}) \Dmu \\ 
T^{\prime \diss} &= \a_1 W + \a_5 (Q + v) \Dmu 
        \\ 
        \T{T}^{\prime \diss}_{(ij)}  &= \a_6 \T{W}_{(ij)} + \a_9 (\T{Q}_{ij} + \T{v}_{(ij)}) \Dmu \\ 
\ratesym^\diss &= \a_5 QW + \a_9 \T{Q}_{ij}\T{W}_{ij} + \a_{12} \tau A + \a_{14} \T{\tau}_{ij} \T{A}_{ij}\\
        &+ \a_15 \tau \Dmu 
    \end{split}
    \label{eq:dissipative_fluxes}
\end{equation}
\emph{\textbf{Reactive fluxes:}}
The reactive Onsager matrix is antisymmetric and couples forces and fluxes such that the resulting product of the force and the corresponding matrix element has the opposite time-reversal signature to that of the flux. This leads to the corresponding constitutive equations,
\begin{equation}
    \begin{split}
        J^{\prime \react} &= - \l_2 W + \l_9 (Q + v) \Dmu 
        \\ 
        \T{J}^{\prime \react}_{(ij)} &= - \l_6 \T{W}_{(ij)} + \l_{10} (\T{Q}_{ij} + \T{v}_{(ij)}) \Dmu \\ 
T^{\prime \react} &= \l_2 A + \l_4 (\rho + \tau) \Dmu  
        \\
        \T{T}^{\prime \react}_{(ij)} &= \l_6 \T{A}_{(ij)} + \l_7 (\T{q}_{ij} + \T{R}_{ij}) \Dmu \\ 
\ratesym^\react &= - \l_4 \tau W - \l_7 \T{\tau}_{ij} \T{W}_{ij} - \l_9 \Omega A - \l_{10} \T{W}_{ij} \T{A}_{ij},
    \end{split}
    \label{eq:reactive_fluxes}
\end{equation}
Finally, the constitutive relation for the flux is
$J'_{ij} =
\left(\frac{1}{d} J^{\prime \diss} \delta_{ij} + \tilde{J}^{\prime \diss}_{ij}\right)
+
\left(\frac{1}{d} J^{\prime \react} \delta_{ij} + \tilde{J}^{\prime \react}_{ij}\right)$
where the explicit expressions are provided in
\eqn{eq:dissipative_fluxes} and \eqn{eq:reactive_fluxes}. Since $J_{(ij)}$ can be expressed in terms of $J'_{(ij)}$ and $D_{ij}$ as
$J_{(ij)} = J'_{(ij)} + D_{(ij)},$ which follows directly from \eqn{eqapp:effective_curr}. Hence, using the leading order form of $D_{(ij)}$ as in \eqn{eqapp:Dij} one can find $J_{ij}$, substituting $J_{ij}$ into \eqn{eq:conteqn} of the main text yields
\begin{equation}
    \begin{split}
        \p_t\rho + &\a\Dmu\p^2\rho + \ub_1 \p^4 \mu + \ub_2 \p^2\p_l (\rho^{-1} \pi_l) \\
        &+ \ub_3 \Dmu \p_i\p_j (\pi_i \pi_j) + \cdots = 0
    \end{split}
    \label{eqapp:EOM_rho}
\end{equation}\par
where, 
\begin{equation}
    \begin{split}
    \a = \frac{\a_{12}}{d},~&\ub_1 = \frac{\a_{10}+\a_{13}(d-1)}{d},~\ub_3 = \a_{14} \\ 
    \ub_2 &= \frac{\Dmu}{d}(\l_9 + \l_{10}(d-1))
    \end{split}
\end{equation}
Similarly, upon explicitly deriving the stress tensor $T_{ij}$ and substituting it into \eqn{eq:conteqn} of the main text, we obtain
\begin{align}
\p_t\pi_i -& \p_i P + \varsigma_1 \p_i\p^2\mu - \varsigma_2 \Dmu \p_i\p_l(\rho^{-1}\pi_l) + \ell_1 \Dmu \p_i \rho \notag\\
         - &\varsigma_3 \Dmu \p^2 (\rho^{-1}\pi_i) + \ell_2 \Dmu \p_j (\pi_i\pi_j) + \cdots = 0
    \label{eqapp:EOM_pi}
\end{align}
where,
\begin{equation}
    \begin{split}
        \varsigma_1 = \frac{1}{d}(\l_2 - \l_6 + d),&~ \varsigma_2 = \frac{1}{2d}(2\a_5 + \a_9 (d-2)) \\ 
        \varsigma_3 = - \frac{\a_9}{2},~\ell_1& = \frac{\l_4}{d},~\ell_2 = \l_7 
    \end{split}
    \nonumber
\end{equation}
The ellipsis denotes irrelevant contributions in the renormalisation group sense.

Equations \eqn{eqapp:EOM_rho} and \eqn{eqapp:EOM_pi} are promoted to stochastic hydrodynamics by incorporating Gaussian white noise terms. {In an equilibrium system, the noise covariance is related to the dissipation matrix and the temperature through the Fluctuation–Dissipation Theorem\cite{chaikin1995principles}. Since our system is active, it is not subject to this constraint \cite{toner1995long}. Here, the only requirement is that it must respect the conservation laws.} These noise contributions are constructed to preserve dipole and momentum conservation, reflecting the intrinsic fluctuations arising in hydrodynamic theory, leading to
\begin{equation}
\begin{split}
    \p_t\rho + \a\Dmu\p^2\rho + \ub_1 \p^4 \mu + \ub_2 \p^2\p_l (\rho^{-1} \pi_l) \\
    + \ub_3 \Dmu \p_i\p_j (\pi_i \pi_j) + \cdots = \p_i\p_j\psi_{ij}(\mbf{r}, t)
\end{split}
\label{eq:EOM_rho1}
\end{equation}
and 
\begin{equation}
\begin{split}
    &\p_t\pi_i - \p_i P + \varsigma_1 \p_i\p^2\mu - \varsigma_2 \Dmu \p_i\p_l(\rho^{-1}\pi_l) \\
    &+ \ell_1 \Dmu \p_i \rho - \varsigma_3 \Dmu \p^2 (\rho^{-1}\pi_i) + \ell_2 \Dmu \p_j (\pi_i\pi_j)\\
    &+ \cdots = \p_j \eta_{ij}(\mbf{r}, t)
\end{split}
\label{eq:EOM_pi1}
\end{equation}
where, 
$\eta_{ij}(\mbf{x}, t)$ and $\psi_{ij}(\mbf{x}, t)$  are Gaussian white noises with $\mean{\eta_{ij}(\mbf{x}, t)} = 0 = \mean{\psi_{ij}(\mbf{x}, t)}$ , and variances given by
\begin{equation}
    \begin{split}
        &\mean{\psi_{ij}(\mbf{x}, t) \psi_{lm}(\mbf{x}^\prime, t^\prime)} \\
        &= d_1 \d(\mbf{x} - \mbf{x}^\prime) \d(t - t^\prime)
        \PB{d_2(\d_{il}\d_{jm} + \d_{im}\d_{jl}) - d_3 \d_{ij}\d_{lm} }
    \end{split}
    \label{eq:noise_rho}
\end{equation}

\begin{equation}
    \begin{split}
        &\mean{\eta_{ij}(\mbf{x}, t) \eta_{lm}(\mbf{x}^\prime, t^\prime)} \\ 
        &= c_1 \d(\mbf{x} - \mbf{x}^\prime) \d(t - t^\prime) \PB{c_2(\d_{il}\d_{jm} + \d_{im}\d_{jl}) - c_3 \d_{ij}\d_{lm} }
    \end{split}
    \label{eq:noise_p}
\end{equation}
where, $c_1, d_1, \cdots$ are constants. Note that our system is out of equilibrium, therefore noise variance is free from following a fluctuation dissipation theorem. The above tensorial form of the noise covariance is dictated by the fact that it has to be symmetric seperately under $i\leftrightarrow j$ or $l\leftrightarrow m$.

The equations of motion governing the dynamics around the steady state $\rho = \rho_0$ and $\bm{\pi} = \bm{\pi}_0 = 0$ are obtained by expanding the fields 
about this configuration. We parametrise the fluctuations as $\rho = \rho_0 + \delta \rho,~\bm{\pi} = \delta \bm{\pi},$ where $\rho_0$ denotes the uniform steady-state density, and $\delta \rho$ and $\delta \bm{\pi}$ represent small fluctuations ($\delta \rho \ll 1$, $\delta \bm{\pi} \ll 1$). Any function of $\rho$ and $\bm{\pi}$ can then be expanded around the steady state. For example, a scalar function $\mu(\rho)$ admits the Taylor expansion
\begin{equation}
    \mu(\rho) = \mu(\rho_0) + \mu'(\rho_0)\,\delta\rho + \mathcal{O}(\delta\rho^2).
\end{equation}
Following similar expansion of \eqn{eq:EOM_rho1} and \eqn{eq:EOM_pi1} results in
\begin{align}
    &\p_t\d\r + \a\Dmu \p^2\d\r + \b_1 \p^4\d\r + \b_2 \p^2\p_j(\d\pi_j) \nonumber\\ 
    &- \bar{\b}_2 \p^2\p_l(\d\r \d\pi_l) + \ub_3 \Dmu\p_i\p_j(\d\pi_i \d\pi_j) + \cdots = \p_i\p_j\psi_{ij}(\bm{r},t) \label{eq:seom_fluc_rho} \\ 
&\p_t\d\pi_i + \l \p_i\d\r + \z_1 \Dmu \p_i\p^2\d\r - \z_2 \Dmu \p_i\p_j\d\pi_j  \nonumber \\
    &- \bar{\z_{}}_2 \Dmu \p_i\p_j(\d\r \d\pi_j
    ) - \z_3 \Dmu \p^2 \d\pi_i + \ell_2 \Dmu \p_j(\d\pi_i \d\pi_j) \nonumber\\
    &+ \cdots = \p_j\eta_{ij}(\bm{r},t) \label{eq:seom_fluc_pi}
\end{align}
Where,
\begin{equation}
    \begin{split}
    \a = \a\r_0,~\b_1 = \ub_1 \mu^\prime(\r_0)&,~\b_2 = \ub_2 \r_0^{-1},~\bar{\b}_2 = \ub_2\r_0^ {-2} \\ 
    \l = \ell_1 \r_0 - P^\prime(\r_0),~\z_1 &= \varsigma_1 \mu^\prime(\r_0),~\z_2 = \varsigma_2 \r_0^{-1}\\
    \bar{\z_{2}} = \varsigma_2 \r_0^{-2}&,~\z_3 = \varsigma_3 \r_0^{-1}
    \end{split}
    \nonumber
\end{equation}
\section{The linear theory}\label{app:lin_th}
We write the linear equations of motion by retaining terms up to first order in the fluctuations in Eqs.~\eqref{eq:seom_fluc_rho} and \eqref{eq:seom_fluc_pi}. This yields the following linearised equations:
\begin{align}
    \p_t\d\r + \a\Dmu \p^2\d\r + \b_1 \p^4\d\r + \b_2 \p^2\p_j\pi_j &= 0 \label{eqapp:lin_eom_rho}\\ 
    \p_t\d\pi_i + \l \p_i\d\r + \z_1 \Dmu \p_i\p^2\d\r - \z_2 \Dmu \p_i\p_j\d\pi_j \nonumber \\
    - \z_3 \Dmu \p^2 \d\pi_i &= 0. \label{eqapp:lin_eom_pi}
\end{align}
Upon Fourier transforming Eqs.~\eqref{eqapp:lin_eom_rho} and \eqref{eqapp:lin_eom_pi}, we obtain
\begin{align}
        - i \o\d\rho - \a\Dmu k^2 \d\rho &+ \b_1 k^4 \d\r - i \b_2 k^2 k_i \d\pi_i = 0 \label{eqapp:lineq_eom_kspace_r} \\
        - i \o\d\pi_i - i \z_1 k_i k^2 \d\r_{} &+ \z_2 \Dmu k_i k_j \d \pi_j \nonumber \\
        &+ \z_3 \Dmu k^2 \d \pi_i + i \l k_i \d\r = 0 \label{eqapp:lineq_eom_kspace_pi}
\end{align}
Moving forward, we drop the $\beta_1$ term as it only produces higher order terms.
Projecting \eqn{eqapp:lineq_eom_kspace_pi} parallel and perpendicular to $\hat{\mathbf{k}}$ decouples the longitudinal and transverse components of $\delta \boldsymbol{\pi}$. Defining $\delta \pi^{L} =  \hat{k}_j \delta \pi_j$ and $\delta \pi_i^{T} = \left( \delta_{ij} - \hat{k}_i \hat{k}_j \right) \delta \pi_j$ the longitudinal mode couples to $\delta \rho$, while the transverse mode evolves independently.
\subsection{Dispersion relations}\label{app:lin_sectors}
\emph{Longitudinal sector:}\label{app:longi_sector}
The resulting coupled equations of motion for $\delta\rho$ and $\d \pi^L$ is written as,
\begin{equation}
    (i\omega\mathbf{I}+D) X = 0 
    \label{eq:lin_mat}
\end{equation}
where the dynamical matrix $D$ is given by,
\begin{equation}
    \begin{split}
        &D = \begin{pmatrix}
                \a\Dmu k^2 & i \b_2 k^3 \\ 
                -i \l k &  -\z \Dmu k^2
            \end{pmatrix}
        ,~
        X = \begin{pmatrix}
                \d\rho \\ 
                \d \pi^L
            \end{pmatrix} \\
    &\z = \z_2 + \z_3
    \end{split} 
    \label{eq:longi_mateq}
\end{equation}
Dispersion relation can be derived by solving the characteristic equation $\det(i\omega\mathbf{I}+D) = 0$ for the angular frequency $\o$ leads to 
\begin{equation}
    \begin{split}
        \o_{\pm} = \f{1}{2}\PB{- i\G_1 \pm  \sqrt{-\G_2}} k^2 \\
        \G_1 = \Dmu(\z - \a)~,~  
        \G_2 = 4\l\b_2 + (\Dmu(\z + \a))^2,
    \end{split}
    \label{eqapp:disp_longi}
\end{equation}
A necessary condition for the steady state to be stable is $\G_1>0$. We now distinguish two cases depending on the sign of $\G_2$. For  $\G_2>0$, stability requires $\G_1>\sqrt{|\G_2|}$.
 When the stability condition is satisfied,
 \begin{equation}
    \begin{split}
        \o_{\pm}  = - i \G_{\mp} k^2
    \end{split}
    \label{eq:disp_diff}
\end{equation} 
where $\G_{\pm} = (\G_1 \pm \sqrt{|\G_2|})$. This implies purely diffusive dynamics. For $\G_2<0$,
the dispersion relation takes the form
\begin{equation}
    \begin{split}
        \o_{\pm} =  \frac{1}{2}(- i &\G_1 \pm \sqrt{|\G_2|}) k^2
    \end{split}
    \label{eq:disp_sound}
\end{equation}
which implies sound mode.

\emph{Transverse sector:}
Application of the projector $\d_{ij} - \hat{k}_i\hat{k}_j$ on to \eqn{eqapp:lineq_eom_kspace_pi} leads to an autonomous equation for $\d\pi^T$ 
\begin{equation}
    \begin{split}
        - i \o\d \pi_i^T + \z_3 \Dmu k^2 \d \pi_i^T = 0
    \end{split}
    \label{eq:FTeom_tr}
\end{equation}
whose dispersion relation reads $\o_{T} = - i \Dmu \z_3 k^2$.

\subsection{Exceptional point}
\label{app:excep}
Here, we investigate the occurrence of an exceptional point in the dynamical matrix $D$ of the longitudinal sector of the system. Exceptional points arise in non-Hermitian systems when two or more eigenvalues and their corresponding eigenvectors coalesce\cite{hanai2020critical}.
Eigenvalues and eigenvectors of the dynamical matrix $D$ are 
\begin{equation}
    \begin{split}
        \vartheta_{\pm} &= -\f{k^2 \G_{\pm}}{2} \\
        \ket{\vartheta_{\pm}^{R}} =     \begin{pmatrix}
                                            \f{i k}{2 \bar{\l_{}} } \G^\prime_{\mp} \\ 
                                            1
                                        \end{pmatrix} ~&,~
        \ket{\vartheta_{\pm}^{L}} =     \begin{pmatrix}
                                            -\frac{i}{2 k \b_2} \G^\prime_{\mp} \\ 
                                            1 
                                        \end{pmatrix}
    \end{split}
\end{equation}
where, $\Gamma^{\prime}_{\pm} = \Dmu (\a+\z) \pm \sqrt{\G_2}$.
Since $D$ is a non-Hermitian matrix, the left and right eigenvectors are distinct. 
Here, $\ket{\vartheta_{\pm}^R}$ and $\ket{\vartheta_{\pm}^L}$ denote the right and left eigenvectors, respectively. These eigenvectors, together with their corresponding eigenvalues, coalesce at $\Gamma_2 = 0$, where $\G^\prime_{+} = \G^\prime_{-}$.
Furthermore, for the right eigenvectors,
\begin{equation}
\tan\theta_{+}=  \f{i k}{2 \l_{} } \G^\prime_{-} \quad  \tan\theta_{-}=  \f{i k}{2 \l_{} } \G^\prime_{+}
\end{equation}
\begin{equation}
\tan(\theta_{+}-\theta_{-})=\frac{ik2 \l_{} (\G^\prime_{-}-\G^\prime_{+})}{{(2 \l_{} })^2-k^2\G^\prime_{-}\G^\prime_{+}}
\end{equation}
As we are working in the hydrodynamic limit i.e. $k\rightarrow 0$, therefore $\tan(\theta_{+}-\theta_{-})\propto -i\sqrt{\G_2}k$, which is our sound speed. Doing the same calculations of the left eigenvector will give $\tan(\theta_{+}-\theta_{-})\propto i\sqrt{\G_2}k$, which is speed for the opposite direction.
\subsection{Correlator}\label{app:correlators}
To find the correlation of $\d\rho$ and $\d\pi^L$ in the linear theory, we use stochastic version of \eqn{eq:lin_mat}, which reads

\begin{align}
    - i \o \d\rho - \a\Dmu k^2 \d\rho - i \b_2 k^3 \d \pi^L = \T{\psi} \label{eq:FTSeom_r} \\
    - i \o \d \pi^L + \z\Dmu k^2 \d \pi^L + i \l k \d \r = \eta^L \label{eq:FTSeom_pi}
\end{align}
Here, $\tilde{\psi}(\mathbf{k},\omega) = -k_i k_j \psi_{ij}(\mathbf{k},\omega)$ and $\eta^{L}(\mathbf{k},\omega)$ denote the longitudinally projected component of $\eta(\mathbf{k},\omega)$, defined as  and $\eta^{L}(\mathbf{k},\omega) = i \hat{k}_i k_j \eta_{ij}(\mathbf{k},\omega)$. The corresponding two-point correlation functions in k-space are then given by
\begin{equation}
    \begin{split}
        &\mean{\T{\psi}(\mbf{k}, \o) \T{\psi}(\mbf{k}^\prime, \o^\prime)} \\
        &= (2\pi) d_1 \d(\mbf{k} + \mbf{k}^\prime) \d(\o + \o^\prime)
         \PB{2d_2 (k_i k_i^\prime)^2 - d_3 k^2 k^{2\prime}}
    \end{split}
\end{equation}
\begin{equation}
    \begin{split}
        &\mean{\eta^L(\mbf{k}, \o) \eta^L(\mbf{k}^\prime,\o^\prime)} \\
        &= - (2\pi) k k^\prime c_1 \d(k+k^\prime) \d(\o+\o^\prime) \PB{2c_2 (\hat{k}_i \hat{k}_i^\prime)^2 - c_3}
    \end{split}
\end{equation}
For the purposes of the present analysis, we are interested in the equal-time correlation functions, defined as
$$
\mean{\d g(k,t)\d g(-k,t)} = \int \mathrm{d}\o \mean{\d g(k,\o) \d g(-k,-\o)}, 
$$
Where $g$ can be any density field ($\r$ or $\bpi$) of our system. The density correlator is given by
\begin{equation}
    \begin{split}
        \mean{\d\r(k,t)\d\r(-k,t)} 
        =
\mathbb{D}\PB{\pi \frac{(\z^2\Dmu^2 + \b_2^2 \l) + \G_2 + \G_1^2}{2\G_1 (\G_2 + \G_1^2)}} k^2
\end{split}
\end{equation}
where, $\mathbb{D} = d_1 (2 d_2 - d_3)$. An important point to note is that the density fluctuations, $\delta \rho^2$, vanish as $k^2$ in the limit $k \to 0$. In an equilibrium system, this behavior would correspond to a compressibility that also vanishes as $k^2$ as $k \to 0$.
In real-space terms, this is equivalent to stating that the compressibility of a system with linear size $L$ scales as $1/L^2$ in the thermodynamic limit $L \to \infty$. The root-mean-square (rms) number fluctuations, $\sqrt{\delta N^2}$, within a volume $V$ scale as $\sqrt{S(k \to 0)\, V}$, where $S(k)$ is the structure factor. This implies
\begin{equation}
\sqrt{\delta N^2} \propto \sqrt{\frac{V}{L^2}} 
\propto L^{\frac{d}{2}-1} 
\propto N^{\frac{1}{2}-\frac{1}{d}},
\end{equation}
where, in the last step, we have used the relation $N \propto V \propto L^d$ \cite{ramaswamy2003active}.
For a three-dimensional system, this scaling implies that the rms number fluctuations grow as $N^{1/6}$, which is significantly smaller than the equilibrium scaling $N^{1/2}$. This phenomenon is known as \emph{hyperuniformity}. It commonly arises in systems with dipole conservation \cite{hexner2017noise} and has also been studied recently in active systems \cite{zheng2024universal,kuroda2023microscopic}.\\ \\
Equal time $\dpi^L$ correlator is given by,
\begin{equation}
    \begin{split}
        \mean{\dpi^L(k,t)\dpi^L(-k,t)}
        =
C\pi \PB{\frac{(\a^2\Dmu^2 + \l_{}) + \G_2 + \G_1^2}{2\G_1 (\G_2 + \G_1^2)}}
\end{split}
\end{equation} 
where, $C = c_1 (2 c_2 - c_3)$.

\emph{Transverse sector}
To calculate the correlator of $\d\bpi^T$, we use stochastic version of \eqn{eq:FTeom_tr}, which reads
\begin{equation}
    \begin{split}
        - i \o \r_0 \d \pi_i^T + \z_3 \Dmu k^2 \d \pi_i^T = \eta_i^T(\mbf{k}, \o)
    \end{split}
    \label{eq:trans_res}
\end{equation}
where, the transeverse projected noise $\eta_i^T(\mbf{k}, \o) = i P_{ij} k_l \eta_{lj}(\mbf{k}, \o) ~;~ P_{ij} = \d_{ij} - \hat{k}_i \hat{k}_j$,
whose variance is given by,
\begin{equation}
    \begin{split}
        \Mean{\eta_i^T(\mbf{k}, \o)& \eta_j^T(\mbf{k}^\prime, \o^\prime)} 
        = - 2 \pi c_1 k_m k_r^\prime \d(k + k^\prime) \d(\o + \o^\prime) \\
        &P_{il} P_{jq}^\prime (c_2(\d_{lr}\d_{mq} + \d_{lq}\d_{mr}) - c_3\d_{lm}\d_{rq})
    \end{split}
    \nonumber
\end{equation}
Hence, the equal time two point correlator takes the form
\begin{equation}
    \begin{split}
        \Mean{\d\pi_i^T&(\mbf{k}, t) \d \pi_j^T(-\mbf{k}, t)} = \frac{2 \pi^2 c_1 c_2}{\r_0 \Dmu \z_3} P_{ij} \propto k^0
    \end{split}
\end{equation}
In passive case this correlation diverges as $1/k^2$ as $k \to 0$ \cite{glorioso2022breakdown} as opposed to active case where it is $k^0$.
\section{Estimation of dynamical exponent in $ d < d_c$}\label{app:scaling}

Under the scaling used in the main paper for \eqn{eq:scaled_eom}, we see from \eqref{eq:noise_rho} and \eqref{eq:noise_p} that the noise $\psi_{ij}$ and $\eta_{ij}$ scale as $ b^{- (z + d)/2}$. After scaling \eqn{eq:EOM_rho1}, (\ref{eq:EOM_pi1}), and
reorganizing to keep the coefficient of the time-derivative
terms, $\partial_t$, at unity \cite{forster1977large}, we get the scaled equations
\begin{equation}
    \begin{split}
&\p_t \d\r  + b^{z-2} \a\Dmu \p^2 \d\r + b^{(z - 3 - \varrho + \Upsilon)} \b_2 \p^2\p_l \d\pi_l +\\
        &b^{(z - \varrho - 2 + 2 \Upsilon)} \ub_3 \Dmu\p_i\p_j( \d\pi_i \d\pi_j) - b^{(z - 3 - \Upsilon)} \bar{\b}_2 \p^2\p_l(\d\r \d\pi_l) \\ 
        &+ b^{z-4} \b_1 \p^4 \d\r + \cdots = b^{(\frac{z}{2} - \varrho - \frac{d}{2} - 2)} \p_i\p_j\psi_{ij}        
    \end{split}
    \label{eqapp:scaled_seom_drho}
\end{equation}
\begin{equation}
    \begin{split}
&\p_t \d\pi + b^{(z - 3 - \Upsilon + \varrho)} \z_1 \Dmu \p_i\p^2 \d\r + b^{(z + \varrho - \Upsilon - 1)} \l \p_i \d\r\\
        &- b^{(z-2)}  \PB{\z_2 \Dmu \p_i\p_l \d\pi_l + \z_3 \Dmu \p^2 \d\pi_i} + b^{(z + \varrho - 2)} \\   
        &\bar{\z_{}}_2 \Dmu \p_i\p_l(\d\r \d\pi_l)
        + b^{(z + \Upsilon - 1)} \ell_2 \Dmu \p_j(\d\pi_i \d\pi_j) \\ 
        &+ \cdots  = b^{(\frac{z}{2} - \Upsilon - \frac{d}{2} - 1)} \p_j \eta_{ij},
    \end{split}
    \label{eqapp:scaled_seom_dpi}
\end{equation}
which were reported in \eqn{eq:scaled_eom} (main text).
For the above set of equations Gaussian fixed point leads to the choice $z = 2, ~\varrho = -(1 + \frac{d}{2}), ~\Upsilon = - \frac{d}{2}$, and nonlinearities {$\p_i\p_j (\dpi_i \dpi_j)$ and} $\p_j(\d\pi_i \d\pi_j)$ are relevant below critical dimension $d_c=2$.

Here, we employ the same heuristic scaling argument introduced in Ref.~\cite{glorioso2022breakdown} to estimate the dynamical critical exponent $z$ in spatial dimensions $d<d_c$. 
As discussed above, for $d<2$ the nonlinear term $\p_j (\d\pi_i \d\pi_j)$ in \eqn{eqapp:scaled_seom_dpi} becomes relevant in the renormalisation–group sense. Under the scaling transformation, time-derivative term and the relevant nonlinear term scales as
$ b^{(\Upsilon - z)} \p_t \d\pi_i \quad \text{and} \quad b^{(2\Upsilon - 1)} \p_j (\d\pi_i \d\pi_j)$ respectively. If we require these two contributions to scale in the same way, we obtain the consistency condition
\begin{equation}
    \Upsilon - z = 2\Upsilon - 1
\end{equation}
which gives $z = 1 - \Upsilon$.
To proceed, we make a crude approximation that, despite the relevance of the nonlinearities for $d<2$, the scaling exponents of  $\d\r$ and $\d\bpi$  remain near their Gaussian fixed-point values. Substituting the Gaussian value of $ \Upsilon$ into the above relation yields 
\begin{equation}
    z = 1 + \frac{d}{2}.
\end{equation}

The same estimate for the dynamical exponent is obtained by comparing the scaling of $b^{\varrho - z}\p_t \d\r$ {and}  $b^{2\Upsilon - 2} \p_i \p_j (\d\pi_i \d\pi_j)$ in \eqn{eqapp:scaled_seom_drho}, again under the assumption that the nonlinear couplings do not feed back into (i.e., renormalise) the Gaussian scaling exponents of $\d\r$ and $\d\bpi$.

\section{Additional details of microscopic lattice model}\label{app:latt_model}
We restate the Hamiltonian for the microscopic model, which we use to verify the hydrodynamic theory in the main text, as follows
\begin{equation}
    \begin{split}
        &H = \sum_{\br} \left[
        \sum_{\mu \nu} \frac{K_\pi}{2}(p^{(\mu)}_{\br} - p^{(\mu)}_{\br + \hat{\nu}} )^2  + \sum_{\mu} V(x^{(\mu)}_{\br} - x^{(\mu)}_{\br + \hat{\mu}}) \right] \\
&V(x) = \sum_n \frac{K_n}{n} x^n.
    \end{split}
    \label{eqapp:latt_model_ddim}
\end{equation}
The displacement $x^{(\mu)}_{\br}$ and momentum $p^{(\mu)}_{\br}$ fields satisfy periodic boundary conditions (PBC) in lattice-site position $\br$, and the meanings for the remaining symbols appearing above are described near \eqn{eq:latt_model_ddim} of the main text. As discussed in \app{app:shift}, the passive free-energy density must be invariant under the transformation $\bpi \to \bpi + \rinv \bm{c}$, where $\bm{c}$ is a constant vector. Analogously, the Hamiltonian in \eqn{eqapp:latt_model_ddim} is invariant under the global shift $p_{\br}^{(\mu)} \to p_{\br}^{(\mu)} + c^{(\mu)}$. This symmetry constrains the kinetic term to depend only on lattice gradients of $p_{\br}^{(\mu)}$. Consequently, the kinetic contribution in \eqn{eqapp:latt_model_ddim} takes the form $\sum_{\br}\sum_{\mu\nu} ({K_\pi}/{2})(p^{(\mu)}_{\br} - p^{(\mu)}_{\br + \hat{\nu}} )^2$. Furthermore, momentum conservation implies invariance under translation, $x_{\br}^{(\mu)} \to x_{\br}^{(\mu)} + a$, where $a$ is a constant. Implying, the potential energy can depend only on relative displacements and a function of the differences $(x_{\br}^{(\mu)} - x_{\br+\hat{\mu}}^{(\mu)})$.
 
For the Hamiltonian in \eqn{eqapp:latt_model_ddim}, these symmetries $(x_{\br}^{(\mu)} \to x_{\br}^{(\mu)} + a,~p^{(\mu)}_{\br} \to p^{(\mu)}_{\br} + c^{(\mu)})$ imply the conservation of the quantities 
\begin{equation}
    D^{(\mu)} = \sum_{\br} \dot{x}_{\br}^{(\mu)},~P^{(\mu)} = \sum_{\br} p^{(\mu)}_{\br}~.
    \label{eqapp:latt_DP}
\end{equation}
These $D^{(\mu)}$, $P^{(\mu)}$ are the microscopic analogue of total dipole and linear momentum introduced in \eqn{eq:conserved_quantities} (main text), respectively.
Conservation of these quantities (\eqn{eqapp:latt_DP}) in the absence of noise can be verified directly using the fundamental Poisson brackets between $x_{\br}^{(\mu)}$ and $p_{\br}^{(\mu)}$.
\begin{equation}
    \CB{x^{(\mu)}_{\br}, p^{(\nu)}_{\br^\prime}} = \d_{\mu,\nu}\d_{\br, \br^\prime}
    \label{eqapp:p_bracket}
\end{equation}
Using poission bracket \eqn{eqapp:p_bracket} one can directly calculate the total time derivative of $D^{(\mu)}$ and $P^{(\mu)}$,
\begin{align}
    \frac{d D^{(\mu)}}{dt} &= \CB{D^{(\mu)}, H} = \sum_{\br} \frac{\d H}{\d p^{(\mu)}_{\br}}\nonumber \\
    &= \sum_{\br} \P{2 p^{(\mu)}_{\br} - p^{(\mu)}_{\br + \hat{\mu}} - p^{(\mu)}_{\br - \hat{\mu}}} = 0 \\ 
    \frac{d P^{(\mu)}}{d t} &= \CB{P^{(\mu)}, H} = - \sum_{\br} \frac{\d H}{\d x^{(\mu)}_{\br}} \\
    &= - \sum_{\br} \PB{V^\prime(x^{(\mu)}_{\br} - x^{(\mu)}_{\br + \hmu}) - V^\prime(x^{(\mu)}_{\br - \hmu} - x^{(\mu)}_{\br})} \nonumber\\
    &= 0. \nonumber
\end{align}
The Hamiltonian dynamics is inherently energy conserving but can be driven out of equilibrium (and made active) by introducing terms that do not arise from the Hamiltonian; in this case, the system does not satisfy the Fluctuation–Dissipation Theorem (FDT). For our lattice model, this is achieved by adding a \emph{local} dissipative term to the dynamics of $p^{(\mu)}_{\br}$ together with a noise term, as follows
\begin{equation}
    \begin{split}
        \dot{p}^{(\mu)}_{\br} &= - \frac{\d H}{\d x^{(\mu)}_{\br}} - \g \dot{x}^{(\mu)}_{\br} - \z^{(\mu)}_{\br} + \z^{(\mu)}_{\br - \hmu} \\ 
        \dot{x}^{(\mu)}_{\br} &= \frac{\d H}{\d p^{(\mu)}_{\br}}, \end{split}
    \label{eqapp:stochastic_EOM}
\end{equation}
where
\begin{equation}
    \begin{split}
        \mean{\z^{(\mu)}_{\br}(t) \z^{(\nu)}_{\br^\prime}(t^\prime)} = 2 T \d_{\mu \nu} \d_{\br \br^\prime} \d(t-t^\prime) ~;~ \mean{\z_{\br}^{(\mu)}} = 0
    \end{split}
    \label{eqapp:noise_corr}
\end{equation}
It can be shown that under the above form of noise and dissipation the total dipole and linear momentum remain conserved, by taking the time derivative of $P^{(\mu)}$, $D^{(\mu)}$ and using \eqn{eqapp:stochastic_EOM} to yield
\begin{equation}
    \frac{d P^{(\mu)}}{d t} = \sum_{\br} \dot{p}^{(\mu)}_{\br} = \sum_{\br} \PB{- \frac{\d H}{\d x^{(\mu)}_{\br}} - \g \dot{x}^{(\mu)}_{\br} - \z^{(\mu)}_{\br} + \z^{(\mu)}_{\br - \hmu}} = 0 
    \label{eqapp:dPdt}
\end{equation}
and
\begin{equation}
        \frac{d D^{(\mu)}}{d t} = \sum_{\br} \dot{x}^{(\mu)}_{\br} = \sum_{\br} \frac{\d H}{\d p^{(\mu)}_{\br}} = 0~.
        \label{eqapp:dDdt}
\end{equation}
To simplify \eqn{eqapp:dPdt}, we have used PBC to show the term $\sum_r (\z_{\br}^{(\mu)} - \z^{(\mu)}_{\br - \hat{\mu}})$ vanishes after summation over lattice-sites $\br$. 
By the same argument, all remaining terms in \eqn{eqapp:dDdt} and \eqn{eqapp:dPdt} also vanish leading to the conservation of the total dipole and momentum.  
\subsection{Continuum limit of lattice model}\label{app:continuum_limit}
Here, we have taken the continuum limit of the discrete Hamiltonian
given in \eqn{eq:latt_model_ddim}. This is done in order to compare and
contrast the resulting continuum theory with the conventional spring–mass
model.\par
The discrete model in \eqn{eq:latt_model_ddim} is defined by setting the lattice spacing to unity. To take the proper continuum limit, we reintroduce the lattice spacing, denoted by $a$, as given below
\begin{equation}
    \begin{split}
        H = \sum_{r} 
        \sum_{\mu \nu} a^d &\frac{(K_\pi a^{2+d})}{2}\frac{(p^{(\mu)}_r - p^{(\mu)}_{r + a\hat{\nu}})^2}{a^{2(1+d)}} \\
        &+ \sum_{r,\mu}\sum_{n=2}^{4} a^d \frac{(K_n a^{n-d})}{n} \frac{(x_r^\mu - x_{r+ a\hat{\mu}}^\mu)^n}{a^{n}}
    \end{split}
    \label{eq:latt_model_with_a}
\end{equation}
Here $\mathbf{r} = (r_1,\ldots,r_d)$, with $r_i = n_i a$, where $n_i$ denotes the
$n_i$-th lattice point in the $i$-th direction. The continuum limit is taken
by letting the lattice spacing $a$ approach zero. Accordingly, $\mathbf{r} \coloneqq \lim_{\substack{a \to 0 \\ n_i \to \infty}} r \qquad \forall\, i,$ where $\mathbf{r}$ becomes a continuous $d$-dimensional vector. Similarly, the continuum limits of the discrete fields $x_r^\mu(t)$ and $p_r^\mu(t)$ are defined as
$\phi^\mu(\mathbf{r},t) \coloneqq \lim_{a \to 0} x_r^\mu(t), ~\pi^\mu(\mathbf{r},t) \coloneqq \lim_{a \to 0} {p_r^\mu(t)}/{a^d}.$ The summation over lattice sites is replaced by a spatial integral according to $
\int \mathrm{d}{\mathbf{r}} \coloneqq \sum a^d.$ The continuum limits of the momentum and spring stiffness parameters are defined as $\kappa_\pi \coloneqq
\lim_{\substack{a \to 0 \\ k_\pi \to \infty}} K_\pi a^{2+d},
~ \kappa_n \coloneqq \lim_{\substack{a \to 0 \\ K_n \to \infty}} K_n a^{n-d}.$ With these identifications, one can take the proper continuum limit ($a \to 0$ and $n_i \to \infty ~\forall~ i$) of Eq.~\eqref{eq:latt_model_with_a}, which results in
\begin{equation}
    H = \int d{\mathbf{r}} \sum_{\mu \nu}\frac{\kappa_\pi}{2} (\p_\nu \pi^\mu)^2 + \sum_\mu \frac{\kappa_n}{n} (\p_\mu \phi^\mu)^n
    \label{eq:conti_lim_latt}
\end{equation}
\postarxiv{This portion needs to be rewritten}
Here, the continuum limits of the fields are identical to those of the nonlinear scalar field theory\cite{peskin2018introduction}. However, the finite-difference structure of the kinetic term leads to a crucial difference. Which fixes the dimension of $\kappa_\pi$, proportional to $[L]^{(2+d)}$. For the quadratic (linear) potential in \eqn{eq:latt_model_with_a}, the coefficient $K_2$ plays the role of the spring stiffness. Which also shares the identical continuum limit with spring constant.
\section{Breakdown of Fluctuation-Dissipation Theorem (FDT)}\label{app:FDT}
We show that the structure of the model, defined by \eqn{eq:latt_model_ddim} and \eqn{eq:stochastic_EOM} of the main text, leads to an explicit breakdown of the fluctuation–dissipation theorem.\\
\noindent
Using the discrete Fourier transforms
\begin{equation}
    x^{(\mu)}_\k = \sum_{\br} e^{i\k \cdot \br} x^{(\mu)}_{\br} ~;~ p^{(\mu)}_\k = \sum_n e^{i \k \cdot \br} p^{(\mu)}_{\br}, 
\end{equation}
\eqn{eq:stochastic_EOM} can be written in the form
\begin{equation}
\label{fdt}
    \dot{\Phi}_{\k}^{(\mu)}
    = - \mathcal{G} F^{(\mu)}_\k + \xi^{(\mu)}_\k,
\end{equation}
where
\begin{equation}
    \begin{split}
    \Phi_\k^{(\mu)} = 
    \begin{pmatrix}
        p^{(\mu)}_\k \\ 
        x_\k^{(\mu)}
    \end{pmatrix}
    ~;~
    F^{(\mu)}_\k &= 
    \begin{pmatrix}
        \frac{\p H}{\p p^{(\mu)}_{-\k}} \\ 
        \frac{\p H}{\p x^{(\mu)}_{-\k}}
    \end{pmatrix}
    ~;~
    D = 
    \begin{pmatrix}
        1 & \g \\ 
        0 & 1 
    \end{pmatrix} \\
    \xi_\k^{(\mu)} = \mathbb{L}_{k_{(\mu)}}
    \begin{pmatrix}
        \z_\k^{(\mu)} \\ 
        \eta_\k^{(\mu)}
\end{pmatrix}
    ~;~ \mathbb{L}_{k_{(\mu)}} &= (e^{-ik_{(\mu)}} - 1) D^{-1} 
    \begin{pmatrix}
        1 & 0 \\
        0 & 0 
    \end{pmatrix}\\
    \mathcal{G} &= i D^{-1} \sigma_y 
    \end{split}
\end{equation}
with $\sigma_y$ being the $y$-component Pauli matrix.
\Eqn{fdt} is written in the language of fluxes, $\dot{\Phi}_{\k}^{(\mu)}$, connected by the forces $F^{(\mu)}_\k$ through an Onsager-Casimir matrix $\mathcal{G}$ \cite{casimir1945onsager}. In this form, the FDT requires the noise covariance matrix to be proportional to the symmetric part of the Onsager-Casimir matrix where the proportnality constant is identified as the temperature \cite{chaikin1995principles,onsager1931reciprocal,casimir1945onsager,dadhichi2018origins}.
Therefore, for our system the condition for FDT to hold would be 
$\mathbb{L}^T_{-\k}\mathbb{L}_{-\k}=\mathcal{G} + \mathcal{G}^T$, 
which clearly cannot be satisfied, since
\begin{equation}
\mathbb{L}^T_{-k_{(\mu)}} \mathbb{L}_{k_{(\mu)}} = \sin^2{\PB{\frac{k_{(\mu)}}{2}}} 
        \begin{pmatrix}
            1 & 0 \\ 
            0 & 0
        \end{pmatrix}
\end{equation}
and
\begin{equation}
 \mathcal{G} + \mathcal{G}^T = 
        \begin{pmatrix}
            \g & 1 \\
            1 & 0
        \end{pmatrix}.
\end{equation}
In other words, the noise covariance matrix is not proportional to the symmetric part of Onsager-Casimir matrix, unlike the model in \rf{glorioso2022breakdown} which uses a FDT-compatible dissipative term, thus making our model intrinsically out-of-equilibrium at the microscopic level.
\section{Quantitative analysis of collapse}\label{app:collapse_err}
In the hydrodynamic limit equal the time momentum correlator $C_\pi(k_x,t) = \mean{p^{(x)}_{-k_x,-k_{\perp}}(t) p^{(x)}_{k_x,k_{\perp}}(t)}|_{k_\perp=0}$ exhibit a universal scaling form 
\begin{equation}
    t^{-\a}C_\pi(k_x/t^{1/z},t)=F(k_x/t^{1/z})
    \label{eqapp:universal_scaling}
\end{equation}
as discussed in the \emph{simulations} paragraph of the main text. We have set the component $k_\perp$ transverse to $k_x$ to be \emph{zero} for computational convenience. Determining the value of the dynamical exponent $z$ that yields the optimal data collapse is the primary objective. To quantify the quality of the collapse, we introduce a measure referred to as the collapse error. This measure assigns a numerical value for a given $(\alpha,z)$ with smaller values of the measure indicating a better collapse onto \eqn{eqapp:universal_scaling}.

Since we aim to collapse $k_x$ vs $C_\pi(k_x,t)$ datasets across large time scales, naturally the scaled wavenumber $\T{k} = t^{-1/z} k_x$ will span several orders of magnitude. To ensure that the scaled correlator $t^{-\alpha} C_\pi(\tilde{k},t)$ contains data points within a common interval, we apply the collapse measure in the window $\tilde{k} \in [\T{k}_{\mathrm{min}}, \T{k}_{\mathrm{max}}]$, where
\begin{align}
    \T{k}_{\text{min}} = \text{max}(k_x^{\text{min}} t^{1/z}),~\T{k}_{\text{max}} = \text{min}(k_x^{\text{max}} t^{1/z})\notag
\end{align}
with $k_x^{\text{min}}$ ($k_x^{\text{max}}$) being the minimum (maximum) $k_x$ value over all datasets.
Next, for each time slice $t_i$, we construct a linearly interpolated function $\mathcal{F}(\tilde{k},t_i)$ from the rescaled correlator $t_i^{-\alpha} C_\pi(\tilde{k},t_i)$. The interpolation is performed over the common interval $\tilde{k} \in [\T{k}_{\mathrm{min}}, \T{k}_{\mathrm{max}}]$. The collapse error is then defined in terms of these interpolated functions
\begin{equation}
    \mathrm{err}(\a,1/z) = \frac{1}{N^2_t N_{\T{k}}} \sum_{\T{k}}\sum_{t_i,t_j} \left[\frac{(\mathcal{F}^(\T{k}, t_i) - \mathcal{F}(\T{k}, t_j))}{\mathrm{max}(t_i,t_j)^\a}\right]^2,
\end{equation}
where $N_{\T{k}}$ is number of $\T{k}$ points in the the interval $\T{k} \in [\T{k}_{\mathrm{min}}, \T{k}_{\mathrm{max}}]$ and $N_t$ is the total number of time steps. The above collapse error was numerically minimized by searching for optimal values in the $(\alpha,1/z)$ plane to obtain the dynamical exponents reported in \fig{fig:collapse} (main text).
\section{Extracting dynamical exponent $z$ from feature tracking}\label{app:tvsk}
\begin{figure}
    \centering
\includegraphics[width=\columnwidth]{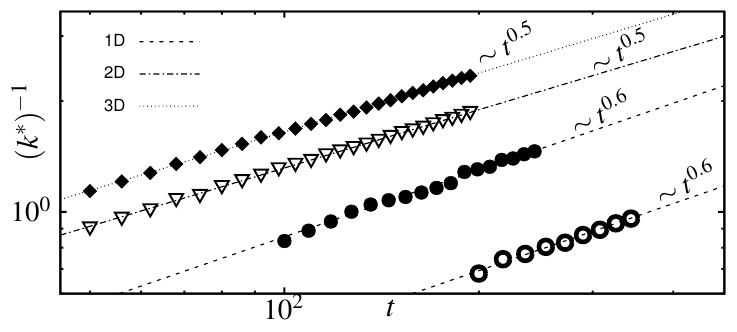}
    \caption{The inverse characteristic wavenumber $1/k^*$ versus $t$ plotted in log-log scale. Using a linear fit, the dynamical exponent $z$, for $d=1, 2$, and $3$ dimensions, can be found from the inverse of the slope obtained from the fit.
    }
    \label{fig:tvsk}
\end{figure}
We adopt an alternative approach \cite{glorioso2022breakdown} to extract the dynamical exponent $z$ by examining the relaxation of $C_\pi(k_x,t)=\mean{p_{-k_x,-k_\perp}(t)\,p_{k_x,k_\perp}(t)}|_{k_\perp=0}$.
Higher wavenumber modes relax faster than lower ones, leading to the emergence of a characteristic feature, such as a peak or knee in the $k_x$ vs $C_\pi(k_x,t)$ plot, which we marked with symbols $\bsqr,~\bm{\triangledown},~\circ,~\bullet$ in \fig{fig:collapse} (main text). Considering the imaginary part of the dispersion to scale as $\mathrm{Im}(\omega)\propto k^{z}$, a mode with wavenumber $k$ relaxes with a timescale $k^{-z}$, implying that the wavenumber $k^*$ of a characteristic feature scales as $k^* \propto t^{-1/z}$. Therefore, tracking the time evolution of $k^*$ provides another method to determine $z$. 
Doing so, we find that $k^*$ indeed scales as a powerlaw in $t$ as shown in \fig{fig:tvsk}. Determinig the \emph{inverse} of the slopes from the log-log plot of the powerlaws gives us the dynamical exponent $z$ in $d=1,2,3$ dimensions, and their values agree with those reported in \fig{fig:collapse} (main text).

\clearpage{}
\end{appendix}
\bibliographystyle{unsrt}
\bibliography{ref} \end{document}